\documentclass{ieeeaccess}
\usepackage{cite}
\usepackage{amsmath,amssymb,amsfonts}
\usepackage{graphicx}
\usepackage{textcomp}
\usepackage{algorithm}
\usepackage{algpseudocode}
\usepackage{float}
\usepackage{mathrsfs}
\usepackage{amsthm}

\usepackage{stfloats}
\usepackage{bm}
\usepackage{subfigure}
\usepackage{color}

\newtheorem{theorem}{Theorem}

\makeatletter
\newenvironment{breakablealgorithm}
{
\begin{center}
\refstepcounter{algorithm}
\hrule height.8pt depth0pt \kern2pt
     \renewcommand{\caption}[2][\relax]{
{\raggedright\textbf{\ALG@name~\thealgorithm} ##2\par}%
\ifx\relax##1\relax 
\addcontentsline{loa}{algorithm}{\protect\numberline{\thealgorithm}##2}%
\else 
\addcontentsline{loa}{algorithm}{\protect\numberline{\thealgorithm}##1}%
\fi
\kern2pt\hrule\kern2pt
}
}{
\kern2pt\hrule\relax
\end{center}
}
\makeatother

\makeatletter
\newcommand{\biggg}{\bBigg@{3}}
\def\bigggl{\mathopen\biggg}

\def\bigggr{\mathclose\biggg}

\def\BibTeX{{\rm B\kern-.05em{\sc i\kern-.025em b}\kern-.08em
    T\kern-.1667em\lower.7ex\hbox{E}\kern-.125emX}}

\begin{document}
\history{Date of publication xxxx 00, 0000, date of current version xxxx 00, 0000.}
\doi{10.1109/ACCESS.2020.2973268}

\title{Performance of High-Mobility MIMO Communications with Doppler Diversity}
\author{\uppercase{Xiaoyun Hou}\authorrefmark{1},
\uppercase{Jie Ling\authorrefmark{1}, and Dongming Wang\authorrefmark{2}}}
\address[1]{Institute of Signal Processing and Transmission, Nanjing University of Posts and Telecommunications, Nanjing, 210003, P.R.China (email: houxy@njupt.edu.cn)}
\address[2]{National Mobile Communications Research Lab, Southeast University, Nanjing, 210096, P.R.China (email: wangdm@seu.edu.cn)}
\tfootnote{This work was supported by the National Natural Science Foundation of China (61971241).}

\markboth
{X. Hou \headeretal: Performance of High-Mobility MIMO Communications with Doppler Diversity}
{X. Hou \headeretal: Performance of High-Mobility MIMO Communications with Doppler Diversity}

\corresp{Corresponding author: Xiaoyun Hou (e-mail: houxy@njupt.edu.cn)}

\begin{abstract}
A rapid change of channels in high-speed mobile communications will lead to difficulties in channel estimation and tracking but can also provide Doppler diversity.
In this paper, the performance of a multiple-input multiple-output system with pilot-assisted repetition coding and spatial multiplexing is studied.
With minimum mean square error (MMSE) channel estimation, an equivalent channel model and the corresponding system model are presented.
Based on random matrix theory, asymptotic expressions of the normalized achievable sum rate of the linear receivers, such as the maximal ratio
combining (MRC) receiver, MMSE receiver and MRC-like receiver, are derived.
In addition, according to the symbol error rate of the MRC-like receiver, the maximum normalized Doppler diversity order and the minimum coding gain loss can be achieved when the repetition number
and signal-to-noise ratio tend to infinity, and the corresponding conditions are derived. Based on the theoretical results, the impacts of different system configurations and channel parameters on the system performance are demonstrated.
\end{abstract}


\titlepgskip=-15pt

\maketitle

\section{Introduction}\label{sec:s1}
\PARstart{W}{ith} the popularization of high-speed data services, the demand for high-performance wireless communication on high-speed trains (HSTs) is also 
increasing. However, as a typical application of 4G/5G, the wireless data throughput on an HST is still a short board in cellular communication systems\cite{AiBo_Comm,Wu2017A,Zhao2018A}. For high-mobility communications, the Doppler spread will far exceed the value considered in the design of traditional mid- and low-speed communication systems. Fast changing small-scale fading and a short coherence time will make it difficult to accurately estimate and track channel parameters with pilot signals. An increase in the channel estimation error can, in turn, result in a degradation in the system performance.

Diversity technology is commonly used antifading technology in wireless mobile communications\cite{Wu2017A}. Since the probability that statistically independent channels experience deep fading at the same time is extremely low, the same signal can be transmitted over different channels to achieve diversity gain. In a high-mobility scenario, considering the fast change of the channel and adding redundancy across different time slots, such as with repetition coding, Doppler diversity can be exploited; therefore, the system performance can be improved. In \cite{Sayeed1999Joint}, joint multipath-Doppler diversity with perfect channel state information (CSI) was proposed, and the validity of the Doppler diversity was proven. In high-mobility scenarios, the channel estimation error will not be negligible. For a fast fading channel with a short coherence time, the channel estimation performance could be worse, while higher Doppler diversity could be exploited\cite{Baissas2000Channel}. In \cite{Zhou2015High}, the performance of Doppler diversity in the case of imperfect channel estimation was studied, and the trade-off between the channel estimation error and Doppler diversity was derived. In \cite{Mahamadu2018Fundamental}, the results were further extended to single-input multiple-output systems.

Currently, research on Doppler diversity is mainly focused on a single transmit antenna system. Multiple-input multiple-output (MIMO) technology could fully utilize the spatial resources and then increase the spectral efficiency without additional transmission power. MIMO technology has become a key technology of many wireless communication standards. By using large-scale antenna arrays at the base station, which is also known as a massive MIMO system, the performance could be further improved \cite{Marzetta2010Noncooperative,8738100,Wang2016An}. In \cite{AiBo_JSAC}, a channel model for a millimeter-wave massive MIMO system was presented, and the simulation results showed that the spatial-domain resources could be exploited with massive antennas. Currently, massive MIMO techniques have been adopted by the 5G NR standard. In \cite{Li2015Pilot} and \cite{Li2016Channel}, channel estimation techniques for massive MIMO systems with high mobility were studied. \cite{You2017BDMA} studied how to overcome the Doppler effect with a massive MIMO system in high-mobility scenarios. In \cite{Zemen2019OP}, the combination of a massive MIMO system and general orthogonal precoding was proposed to utilize full diversity in doubly selective channels. Other coding methods, such as the Alamouti code, can also be combined with repetition code to get more performance gains. The application of Doppler diversity technology in an MIMO system could improve the throughput and reliability of high-mobility wireless communications. However, in the case of imperfect CSI, the performance of an MIMO system using Doppler diversity has not been studied.

In this paper, we investigate the performance of high-mobility MIMO communications with Doppler diversity under imperfect CSI. For the convenience of analysis, we adopt simple repetition coding. To improve the spectral efficiency, we consider spatial multiplexing with repetition coding. The major contributions of this paper include:
\begin{enumerate}
\item For an MIMO system with pilot-assisted repetition coding and spatial multiplexing, an equivalent channel model of the MIMO time-varying channel with imperfect channel estimation is established. The model can be regarded as a more general form of \cite{Zhou2015High,Mahamadu2018Fundamental}.
\item For maximum ratio combining (MRC) and MRC-like linear receivers, an asymptotic expression of the SINR of the system under Doppler diversity is derived. The deterministic equivalent of the normalization rate for the minimum mean squared error (MMSE) receiver is presented. The results show that the method can obtain a good approximation of the normalization rate even when the antenna size is small and the number of repetitions is small.
\item The performance of the Doppler diversity, including the diversity order and minimum coding gain loss for an MRC-like receiver, is derived. An explicit relationship between the Doppler diversity gain, coding gain loss and system parameters is revealed.
\end{enumerate}

The rest of this paper is organized as follows: Section II provides the signal model, channel model, channel estimation and equivalent model of an MIMO system using pilot-assisted repetition coding. Section III studies the corresponding SINR performance of three different receivers in the presence of imperfect CSI. Section IV derives the Doppler diversity order and coding gain loss of an MRC-like receiver based on the average symbol error rate (SER). Section V presents the numerical results, and Section VI draws the conclusion of the paper.

The symbols used in this paper are described below. Bold lowercase letters and bold uppercase letters represent vectors and matrices, respectively. ${{\bm{I}}_{M}}$ denotes the unit matrix with dimensions of $M\times M$. $|\cdot |$ represents the absolute value of a scalar. ${\left[  \cdot  \right]^{\text{T}}}$ and ${{\left[ \cdot  \right]}^{\text{H}}}$ represent vector or matrix transposes and conjugate transposes, respectively. ${{\mathcal{R}}^{m\times n}}$ and ${{\mathcal{C}}^{m\times n}}$ represent the set of $m\times n$-dimensional real and complex matrices, respectively. $\text{E}\left[ \cdot  \right]$ and $\operatorname{cov}\left[ \cdot  \right]$ represent mathematical expectation and covariance, respectively. $\text{Tr}\left[ \cdot  \right]$ is the trace of a matrix. $\text{diag}\left( \bm{x} \right)$ represents a diagonal matrix with $\bm{x}$ as the main diagonal value. $\mathcal{C}\mathcal{N}(0,\sigma _{{}}^{2})$ denotes a circular symmetric complex Gaussian (CSCG) distribution with zero mean and variance ${{\sigma }^{2}}$. $\xrightarrow{a.s.}$ denotes almost sure (a.s.) convergence.

\section{System Model}\label{sec:s2}

\subsection{System Description}\label{sec:ss2_1}
Consider an MIMO wireless communications system with ${{N}_{\text{T}}}$ transmit antennas and ${{N}_{\text{R}}}$ receive antennas. Assume that the terminal is moving at high speed, and the maximum Doppler spread is ${{f}_\text{D}}$. At the transmitter, the modulated symbols are converted from series to parallel to produce multiple 
data blocks and are then mapped to different transmit antennas. To obtain the CSI, pilot signals are inserted between duplicate data blocks. To avoid interference between the antennas, different transmit antennas use orthogonal pilots. For convenience, each transmit antenna transmits the pilot signals at different slots.

As shown in Fig.\ref{fig1}, the pilot and data symbols are repeatedly transmitted $N$ times in a frame. Let ${{p}_{i}}$ denote the $i$-th pilot 
symbol and ${{s}_{t,k}}$ denote the $k$-th data symbol transmitted by the $t$-th transmit antenna. The time interval between adjacent pilot slots is ${{T}_{\text{P}}}=({{N}_{\text{T}}}+K)T$, where $T$ is the symbol interval time of the channel. The energies allocated to each pilot symbol and data symbol are ${{E}_{\text{P}}}$ and ${{E}_{\text{C}}}$, respectively. According to proposition 2 in \cite{Ning2014Maximizing}, as long as the pilot interval satisfies $\left( {{N}_{\text{T}}}+K \right)T\le {0.5}/{{{f}_\text{D}}}\;$, an accurate channel estimation can be obtained.

\Figure[!h][width=84mm]{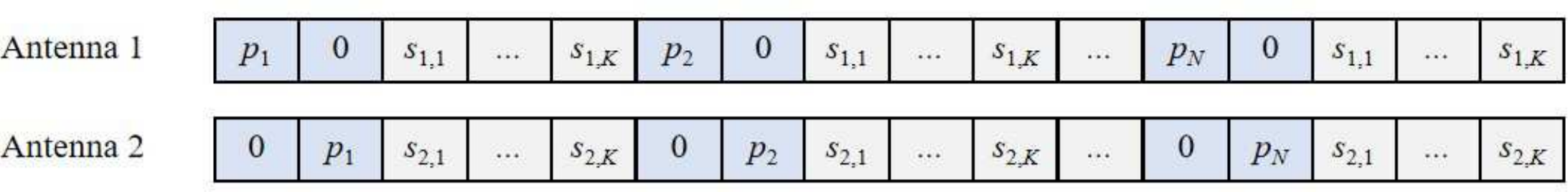}
{Pilot-assisted MIMO system with repetition coding (${{N}_{\text{T}}}$=2)\label{fig1}}

\subsection{Channel Model}\label{sec:ss2_2}
Let ${{h}_{t,r}}(n)$ denote the $n$-th discrete-time channel coefficient between the $t$-th transmit antenna and the $r$-th receive antenna. The channels between different antenna pairs are assumed to be independent, experience wide-sense-stationary uncorrelated scattering (WSSUS) and undergo Rayleigh fading, from \cite{Mahamadu2018Fundamental}
\begin{equation}\label{eq:eq1}
{\rm{E}}\left[ {{h_{{t_1},{r_1}}}\left( m \right)h_{{t_2},{r_2}}^{\rm{*}}\left( n \right)} \right]{\rm{ = }}0,\;\;{r_1} \ne {r_2}\;{\rm{or}}\;\;{t_1} \ne {t_2},
\end{equation}
where ${{h}_{t,r}}(n)$ is a symmetric complex Gaussian process with zero mean; according to \cite{Mahamadu2018Fundamental},
\begin{equation}\label{eq:eq2}
\text{E}\left[ {{h}_{t,r}}\left( m \right)h_{t,r}^{\text{*}}\left( n \right) \right]={{J}_{0}}\left( 2\pi {{f}_{\text{D}}}|m-n|T \right),
\end{equation}
where ${{J}_{0}}\left( \centerdot  \right)$ is the zeroth-order Bessel function of the first kind. Here, it is assumed that ${{f}_{\text{D}}}$ is known, and the estimation of
${{f}_{\text{D}}}$ is described in \cite{Hua2009Novel}.

\subsection{Pilot Signal Model}\label{sec:ss2_3}
Denote the timing indices of the $n$-th pilot symbol from the $t$-th transmit antenna in a transmission frame as:
\[{{i}_{t,n}}=t+(n-1)({{N}_{\text {T}}}+K),n=1,\cdots ,N .\]
According to \cite{Mahamadu2018Fundamental}, the pilot symbols received by the $r$-th receive antenna from the $t$-th transmit antenna can be expressed as:
\begin{equation}\label{eq:eq3}
{{\bm{y}}_{\text{P},t,r}}=\sqrt{{{E}_{\text{P}}}}{{\bm{X}}_{\text{P}}}{{\bm{h}}_{\text{P},t,r}}+{{\bm{z}}_{\text{P},t,r}},
\end{equation}
where
\[{{\bm{y}}_{\text{P},t,r}}={{\left[ \begin{matrix}
   {{y}_{r}}\left( {{i}_{t,1}} \right) & \cdots  & {{y}_{r}}\left( {{i}_{t,N}} \right)  \\
\end{matrix} \right]}^{\text{T}}}\in {{\mathcal{C}}^{N\times 1}} ,\]
\[{{\bm{X}}_{\text{P}}}=\text{diag}\left( \left[ \begin{matrix}
   {{p}_{1}} & \cdots  & {{p}_{N}}  \\
\end{matrix} \right] \right)\in {{\mathcal{C}}^{N\times N}} ,\]
\[{{\bm{h}}_{\text{P},t,r}}={{\left[ \begin{matrix}
   {{h}_{t,r}}\left( {{i}_{t,1}} \right) & \cdots  & {{h}_{t,r}}\left( {{i}_{t,N}} \right)  \\
\end{matrix} \right]}^{\text{T}}}\in {{\mathcal{C}}^{N\times 1}},\]
\[{{\bm{z}}_{\text{P},t,r}}={{\left[ \begin{matrix}
   {{z}_{r}}\left( {{i}_{t,1}} \right) & \cdots  & {{z}_{r}}\left( {{i}_{t,N}} \right)  \\
\end{matrix} \right]}^{\text{T}}}\in {{\mathcal{C}}^{N\times 1}},\]
are the received pilot signal vector, transmitted pilot matrix, channel vector and additive white Gaussian noise (AWGN) vector, respectively. It is assumed that the noise vector is a zero-mean CSCG random vector with covariance matrix ${{\sigma }^{2}}{{\bm{I}}_{N}}$.

\subsection{MMSE Channel Estimation}\label{sec:ss2_4}
The timing indices of the $k$-th data symbol from the $t$-th transmit antenna in a transmission frame can be expressed as:
\[{{k}_{n}}={{N}_{\text {T}}}+k+(n-1)({{N}_{\text {T}}}+K),n=1,\cdots ,N .\]
Then, the vector of all channel coefficients corresponding to the $k$-th data symbol between the $t$-th transmit antenna and the $r$-th receive antenna is
\[{{\bm{h}}_{t,r,k}}={{\left[ \begin{matrix}
   {{h}_{t,r}}\left( {{k}_{1}} \right) & \cdots  & {{h}_{t,r}}\left( {{k}_{N}} \right)  \\
\end{matrix} \right]}^{\text{T}}} .\]
With the received $N$ slot pilot signals and the known channel statistics and according to the principle of the MMSE, the channel estimation of ${{\bm{h}}_{t,r,k}}$ can be expressed as \cite{Zhou2015High},
\begin{equation}\label{eq:eq4}
{{\bm{\hat{h}}}_{t,r,k}}=\sqrt{{{E}_\text{P}}}{{\bm{R}}_{\text{P},t,k}}\bm{X}_{\text{P}}^{\text{H}}{{\left( {{E}_{\text{P}}}{{\bm{X}}_{\text{P}}}{{\bm{R}}_{\text{P}}}\bm{X}_{\text{P}}^{\text{H}}+{{\sigma }^{2}}{{\bm{I}}_{N}} \right)}^{-1}}{{\bm{y}}_{\text{P},t,r}},
\end{equation}
where
\[{{\bm{R}}_{\text{P}}}=\text{E}\left[ {{\bm{h}}_{\text{P},t,r}}\bm{h}_{\text{P},t,r}^{\text{H}} \right]\in {{\mathcal{R}}^{N\times N}},\]
is a Toeplitz matrix, and its first column is given by
\[{{\bm{r}}_{\text{P}}}={{\left[ \begin{matrix}
   {{\rho }_{0}} & \cdots  & {{\rho }_{N-1}}  \\
\end{matrix} \right]}^{\text{T}}},{{\rho }_{n}}={{J}_{0}}\left( 2\pi {{f}_{\text{D}}}\left| n \right|{{T}_{\text{P}}} \right).\]
${{\bm{R}}_{\text{P}}}$ is not related to $t$ and $r$. Because of the symmetry, the first column is the same as the first row. In (\ref{eq:eq4}),
\[{{\bm{R}}_{\text{P},t,k}}\text{=E}\left[ {{\bm{h}}_{t,r,k}}\bm{h}_{\text{P},t,r}^{\text{H}} \right]\in {{\mathcal{R}}^{N\times N}},\]
is also a Toeplitz matrix. Because the zeroth-order Bessel function of the first kind ${{J}_{0}}\left( x \right)$ is even when $x$ is real, the first column and the first row can be written as
\[{{\left[ \begin{matrix}
   {{\tau }_{t,0}} & {{\tau }_{t,-1}} & \cdots  & {{\tau }_{t,-N+1}}  \\
\end{matrix} \right]}^{\text{T}}},{{\left[ \begin{matrix}
   {{\tau }_{t,0}} & {{\tau }_{t,1}} & \cdots  & {{\tau }_{t,N-1}}  \\
\end{matrix} \right]}},\]
where
\[{{\tau }_{t,n}}={{J}_{0}}\left( 2\pi T\left[ {{N}_{\text{T}}}+k-t-n\left( {{N}_{\text{T}}}+K \right) \right] \right),\]
is not related to $r$. When orthogonal pilot sequences are used, that is, ${{\bm{X}}_{\text{P}}}\bm{X}_{\text{P}}^{\text{H}}={{\bm{I}}_{N}}$, by using the following matrix inversion equation,
\[{\bm{CD}}\left( {{\bm{A}} + {\bm{BCD}}} \right) = {\left( {{{\bm{C}}^{ - 1}} + {\bm{D}}{{\bm{C}}^{ - 1}}{\bm{B}}} \right)^{ - 1}}{\bm{D}}{{\bm{A}}^{ - 1}},\]
(\ref{eq:eq4}) can be rewritten as
\begin{equation}\label{eq:eq5}
{{\bm{\hat{h}}}_{t,r,k}}=\sqrt{{{E}_\text{P}}}{{\bm{R}}_{\text{P},t,k}}{{\left( {{E}_{\text{P}}}{{\bm{R}}_{\text{P}}}+{{\sigma }^{2}}{{\bm{I}}_{N}} \right)}^{-1}}\bm{X}_{\text{P}}^{\text{H}}{{\bm{y}}_{\text{P},t,r}}.
\end{equation}
Define the channel estimation error as
\[{{\bm{\tilde{h}}}_{t,r,k}}={{\bm{h}}_{t,r,k}}-{{\bm{\hat{h}}}_{t,r,k}}.\]
Then, according to the MMSE estimation principle, the covariance matrices for the estimated channel vector ${{\bm{\hat{h}}}_{t,r,k}}$ and the error ${{\bm{\tilde{h}}}_{t,r,k}}$ can be expressed as \cite{e21060573}
\begin{equation}\label{eq:eq6}
{{\bm{\hat{R}}}_{t,k}}={{\bm{R}}_{\text{P},t,k}}{{\left( {{\bm{R}}_{\text{P}}}+\frac{1}{{{\gamma }_{\text{P}}}}{{\bm{I}}_{N}} \right)}^{-1}}\bm{R}_{\text{P},t,k}^{\text{H}},
\end{equation}
\begin{equation}\label{eq:eq7}
{{\bm{\tilde{R}}}_{t,k}}={{\bm{R}}_{k}}-{{\bm{R}}_{\text{P},t,k}}{{\left( {{\bm{R}}_{\text{P}}}+\frac{1}{{{\gamma }_{\text{P}}}}{{\bm{I}}_{N}} \right)}^{-1}}\bm{R}_{\text{P},t,k}^{\text{H}},
\end{equation}
respectively, where ${{\gamma }_{\text{P}}}={{{E}_{\text{P}}}}/{{{\sigma }^{2}}}\;$ is the pilot signal-to-noise ratio and ${{\bm{R}}_{k}}=\text{E}\left[ {{\bm{h}}_{t,r,k}}\bm{h}_{t,r,k}^{\text{H}} \right]$. According to the frame structure and the properties of the Bessel function, ${{\bm{R}}_{k}}={{\bm{R}}_{\text{P}}}$.

\subsection{Equivalent System Model}\label{sec:ss2_5}
In this subsection, we establish a signal model for the pilot-assisted repetition precoding MIMO system in the presence of imperfect CSI. Let
$${{\bm{y}}_{k}}={{[\bm{y}_{1,k}^{\text{T}},\cdots ,\bm{y}_{{{N}_{\text{R}}},k}^{\text{T}}]}^{\text{T}}},$$
denote the collection of $N$ slot received signals for all of the receive antennas corresponding to the $k$-th data vector $\bm{s}_k$. ${{\bm{y}}_{k}}$ can be expressed as
\begin{equation}\label{eq:eq8}
{{\bm{y}}_{k}}=\sqrt{{{E}_{\text{C}}}}{{\bm{H}}_{k}}{{\bm{s}}_{k}}+{{\bm{z}}_{k}},
\end{equation}
where the channel matrix ${{\bm{H}}_{k}}$ is denoted as:
\[{{\bm{H}}_{k}}={{\left[ \begin{matrix}
   \bm{H}_{1,k}^{\text{T}} & \cdots  & \bm{H}_{{{N}_{\text{R}}},k}^{\text{T}}  \\
\end{matrix} \right]}^{\text{T}}},\]
\[{{\bm{H}}_{r,k}}=\left[ \begin{matrix}
   {{\bm{h}}_{1,r,k}} & \cdots  & {{\bm{h}}_{t,r,k}} & \cdots  & {{\bm{h}}_{{{N}_{\text{T}}},r,k}}  \\
\end{matrix} \right].\]
According to the definition of the channel estimation error, the channel matrix can be expressed as
\begin{equation}\label{eq:eq9}
{{\bm{H}}_{k}}={{\bm{\hat{H}}}_{k}}+{{\bm{\tilde{H}}}_{k}}.
\end{equation}
Then, (\ref{eq:eq8}) can be rewritten as
\begin{equation}\label{eq:eq10}
{{\bm{y}}_{k}}=\sqrt{{{E}_{\text{C}}}}{{\bm{\hat{H}}}_{k}}{{\bm{s}}_{k}}+{{\bm{\hat{z}}}_{k}},
\end{equation}
where
$${{\bm{\hat{z}}}_{k}}=\sqrt{{{E}_{\text{C}}}}{{\bm{\tilde{H}}}_{k}}{{\bm{s}}_{k}}+{{\bm{z}}_{k}}.$$
Define
$${\bm{\tilde R}_k} \buildrel \Delta \over = {\bm{I}_{{{N_\text{R}}}}} \otimes \left( {\sum\limits_{t = 1}^{{{N_\text{T}}}} {{\bm{\tilde R}_{t,k}}}  + \frac{1}{{{\gamma _{\text{C}}}}}{\bm{I}_N}} \right),$$
where ${{\gamma }_{\text{C}}}={{{E}_{\text{C}}}}/{{{\sigma }^{2}}}\;$. With (\ref{eq:eq7}), we have
\begin{equation}\label{eq:eq11}
{\mathop{\rm cov}} \left( {{{{\bf{\hat z}}}_k},{{{\bf{\hat z}}}_k}} \right) = {E_\text{C}}{{\bm{\tilde R}}_k}.
\end{equation}
Moreover, the covariance matrix of the $t$-th column of ${{\bm{\hat{H}}}_{k}}$ can be expressed as
\begin{equation}\label{eq:eq12}
{\mathop{\rm cov}} \left( {{{{\bm{\hat h}}}_{t,k}},{{{\bm{\hat h}}}_{t,k}}} \right) = {{\bm{I}}_{{{N_\text{R}}}}} \otimes {{\bm{\hat R}}_{t,k}}.
\end{equation}

\section{Capacity Analysis of MRC, MMSE and MRC-like Receivers}\label{sec:s3}
Next, we study the spectral efficiency of the system represented in formula (\ref{eq:eq10}) with different receivers. We first study the MRC receiver and present its asymptotic rate analysis. Then, the deterministic equivalent for the SINR of the MMSE receiver is studied. Finally, the asymptotic rate of MRC-like receivers is analyzed.

\subsection{Asymptotic Achievable Sum Rate of MRC Detection}\label{sec:ss3_1}
When the system uses the MRC receiver, the detection of the $k$-th data symbol from the $t$-th transmit antenna can be expressed as
\begin{align}\label{eq:eq13}
 {\hat{s}}_{\text{MRC},t,k}=&{{E}_{\text{C}}}\bm{\hat{h}}_{t,k}^{\text{H}}{{\bm{\hat{h}}}_{t,k}}{{s}_{t,k}}+{{E}_{\text{C}}}\sum\limits_{l=1,l\ne t}^{{{N}_{\text{T}}}}{\bm{\hat{h}}_{t,k}^{\text{H}}{{{\bm{\hat{h}}}}_{l,k}}{{s}_{l,k}}} \nonumber\\
 &+\sqrt{{{E}_{\text{C}}}}\bm{\hat{h}}_{t,k}^{\text{H}}{{\bm{\hat{z}}}_{k}}.
\end{align}
Therefore, the corresponding SINR can be expressed as \cite{Cao2018Uplink}
\begin{equation}\label{eq:eq14}
{{\gamma }_{\text{MRC},t,k}}=\frac{{{\left| \bm{\hat{h}}_{t,k}^{\text{H}}{{{\bm{\hat{h}}}}_{t,k}} \right|}^{2}}}{\bm{\hat{h}}_{t,k}^{\text{H}}\left( {{{\bm{\hat{H}}}}_{\left[ t \right],k}}\bm{\hat{H}}_{\left[ t \right],k}^{\text{H}}+{{{\bm{\tilde{R}}}}_{k}} \right){{{\bm{\hat{h}}}}_{t,k}}},
\end{equation}
where ${{\bm{\hat{H}}}_{\left[ t \right],k}}$ is the result of ${{\bm{\hat{H}}}_{k}}$ removing the $t$-th column.
\begin{theorem}\label{THEOREM 1}
For the MRC receiver, when $N{{N}_{\text{R}}}\to \infty $, there is
\begin{equation}\label{eq:eq15}
{{\gamma }_{\text{MRC},t,k}}\xrightarrow{a.s.}{{\bar{\gamma }}_{\text{MRC},t,k}},
\end{equation}
where ${{\bar{\gamma }}_{\text{MRC},t,k}}$ is given at the bottom of this page and ${{\gamma }_{\text{C}}}={{{E}_{\text{C}}}}/
{{{\sigma }^{2}}}$ is the data signal-to-noise ratio.
\end{theorem}
\begin{IEEEproof}
See Appendix \ref{sec:Appendix_I}.
\end{IEEEproof}
\begin{figure*}[!b]
\hrulefill \normalsize
\begin{equation}\label{eq:eq16}
\begin{aligned}
{{\bar{\gamma }}_{\text{MRC},t,k}}=\frac{{{N}_{\text{R}}}\text{T}{{\text{r}}^{2}}\left( {{{\bm{\hat{R}}}}_{t,k}} \right)}{\sum\limits_{{{l}_{1}}=1,{{l}_{1}}\ne t}^{{{N}_{\text{T}}}}{\text{Tr}\left( {{{\bm{\hat{R}}}}_{{{l}_{1}},k}}{{{\bm{\hat{R}}}}_{t,k}} \right)}+\text{Tr}\left( {{{\bm{\hat{R}}}}_{t,k}}\left( \sum\limits_{{{l}_{2}}=1}^{{{N}_{\text{T}}}}{{{{\bm{\tilde{R}}}}_{{{l}_{2}},k}}}+\frac{1}{{{\gamma }_{\text{C}}}}{{\bm{I}}_{N}} \right) \right)}.
\end{aligned}
\end{equation}
\end{figure*}
Since repetition coding requires multiple time slots to transmit the same data block, we define the normalized achievable sum rate as follows to better reflect the spectral efficiency of the system:
\begin{equation}\label{eq:eq17}
{{R}_{\text{MRC},k}}=\frac{1}{N}\sum\limits_{t=1}^{{{N}_{\text{T}}}}{{{\log }_{2}}\left( 1+{{\gamma }_{\text{MRC},t,k}} \right)}.
\end{equation}
From Remark 5 of \cite{6172680}, by the control convergence theorem, the following equation is satisfied:
\begin{equation}\label{eq:eq18}
{{R}_{\text{MRC},k}}-\frac{1}{N}\sum\limits_{t=1}^{{{N}_{\text{T}}}}{{{\log }_{2}}\left( 1+{{{\bar{\gamma }}}_{\text{MRC},t,k}} \right)}\xrightarrow{a.s.}0,
\end{equation}
when $N{{N}_{\text{R}}}\to \infty $. According to this asymptotic result, we can obtain an approximate expression of the normalized sum rate of MRC.

\subsection{Asymptotic Achievable Sum Rate of MMSE Detection}\label{sec:ss3_2}
In this section, we analyze the SINR of the MMSE receiver in the presence of imperfect CSI and present a method to calculate the normalized achievable sum rate of the system based on the deterministic equivalent.

When linear MMSE detection is used, the output SINR of the $k$-th symbol from the $t$-th transmit antenna can be expressed as \cite{Li2016Uplink}
\begin{equation}\label{eq:eq19}
\begin{aligned}
{{\gamma }_{\text{MMSE},t,k}}=\bm{\hat{h}}_{t,k}^{\text{H}}{{\left( {{{\bm{\hat{H}}}}_{\left[ t \right],k}}\bm{\hat{H}}_{\left[ t \right],k}^{\text{H}}+{{{\bm{\tilde{R}}}}_{k}} \right)}^{-1}}{{\bm{\hat{h}}}_{t,k}}.
\end{aligned}
\end{equation}
Let
$${{\bm{\breve{h} }}_{t,k}} = \left[ {{{\bm{I}}_{{{N_\text{R}}}}} \otimes {{\left( {\sum\limits_{t = 1}^{{{N_\text{T}}}} {{{{\bm{\tilde R}}}_{t,k}}}  + \frac{1}{{{\gamma _{\text{C}}}}}{{\bm{I}}_N}} \right)}^{ - \frac{1}{2}}}} \right]{{\bm{\hat h}}_{t,k}},
$$
$${{\bm{\breve{H} }}_k}{\rm{ = }}\left[ {\begin{array}{*{20}{c}}
{{{{\bm{\breve{h} }}}_{1,k}}}& \cdots &{{{{\bm{\breve{h} }}}_{{{N_\text{T}}},k}}}
\end{array}} \right].$$
Then, we have
$${\gamma _{{\text{MMSE}},t,k}}{\rm{ = }}{\bm{\breve{h} }}_{t,k}^{\rm{H}}{\left( {{{{\bm{\breve{H} }}}_{\left[ t \right],k}}{\bm{\breve{H} }}_{\left[ t \right],k}^{\rm{H}} + {{\bm{I}}_{N{{N_\text{R}}}}}} \right)^{ - 1}}{{\bm{\breve{h} }}_{t,k}}.
$$
${{\bm{\breve{H}}}_{\left[t\right],k}}$ is a submatrix of ${{\bm{\breve{H}}}_{k}}$ that removes column $t$.
Note that
\begin{align*}
&\operatorname{cov}\left( {{{\bm{\breve{h}}}}_{t,k}},{{{\bm{\breve{h}}}}_{t,k}}\right)={{\bm{I}}_{{{N}_{\text{R}}}}}\otimes\\
&\left({{\left(\sum\limits_{l=1}^{{{N}_{\text{T}}}}{{{{\bm{\tilde{R}}}}_{l,k}}}+\frac{1}{{{\gamma}_{\text{C}}}}{{\bm{I}}_{N}}\right)}^{\text{-}\frac{1}{2}}}{{{\bm{\hat{R}}}}_{t,k}}{{\left(\sum\limits_{l=1}^{{{N}_{\text{T}}}}{{{{\bm{\tilde{R}}}}_{l,k}}}+\frac{1}{{{\gamma}_{\text{C}}}}{{\bm{I}}_{N}}\right)}^{\text{-}\frac{1}{2}}}\right).
\end{align*}
According to Theorem 3.4 of \cite{Couillet2011Random}, when $N{{N}_{\text{R}}}\to \infty $:
\begin{align}\label{eq:eq20}
\frac{1}{N{{N}_{\text{R}}}}{{\gamma }_{\text{MMSE},t,k}}\xrightarrow{a.s.}{{m}_{\mathbf{B},\mathbf{Q},t,k}}.
\end{align}
Define
\[{\bm{B}_{t,k}}\triangleq {{\bm{\breve{H}}}_{\left[t\right],k}}\bm{\breve{H}}_{\left[t\right],k}^{\text{H}},\]
\[{\bm{Q}_{t,k}}\triangleq\operatorname{cov}\left( {{{\bm{\breve{h}}}}_{t,k}},{{{\bm{\breve{h}}}}_{t,k}}\right),\]
\[{{m}_{\bm{B},\bm{Q},t,k}}\triangleq\frac{1}{N{{N}_{\text{R}}}}\text{Tr}\left[{\bm{Q}_{t,k}}{{\left({\bm{B}_{t,k}}+{{\bm{I}}_{N{{N}_{\text{R}}}}}\right)}^{-1}}\right].\]
According to Theorem 1 of \cite{6172680}:
\begin{equation}\label{eq:eq21}
{{m}_{\bm{B},\bm{Q},t,k}}\xrightarrow{a.s.}m_{\bm{B},\bm{Q},t,k}^{\text{o}}.
\end{equation}
In (\ref{eq:eq21}),
\begin{equation}\label{eq:eq22}
m_{\bm{B},\bm{Q},t,k}^{\text{o}}=\frac{1}{N{{N}_{\text{R}}}}\text{Tr}\left( {\bm{Q}_{t,k}}{\bm{T}_{t,k}} \right),
\end{equation}
\begin{equation}\label{eq:eq23}
{\bm{T}_{t,k}}={{\left( \sum\limits_{l=1,l\ne t}^{{{N}_{\text{T}}}}{\frac{{{\bm{\Phi }}_{l,t,k}}}{1+{{e}_{l,t,k}}}}+{{\bm{I}}_{N{{N}_{\text{R}}}}} \right)}^{-1}},
\end{equation}
\begin{equation}\label{eq:eq24}
{{\bm{\Phi }}_{l,t,k}}=\operatorname{cov}\left( {{{\bm{\breve{h}}}}_{l,k}},{{{\bm{\breve{h}}}}_{l,k}}\right),
\end{equation}
where $l$ in (\ref{eq:eq24}) is not equal to $t$ and ${{e}_{l,t,k}}$ is the only solution to the following equation:
\begin{equation}\label{eq:eq25}
{{e}_{l,t,k}}\text{=Tr}\left( {{\bm{\Phi }}_{l,t,k}}{\bm{T}_{t,k}} \right).
\end{equation}
The matrix ${\bm{T}_{t,k}}$ can be obtained by iterative calculations using (\ref{eq:eq23})(\ref{eq:eq25}). The specific algorithm is presented in Appendix \ref{sec:Appendix_II}. Then, the deterministic equivalent of ${{\gamma }_{\text{MMSE},t,k}}$ can be obtained without knowing the exact value of ${{\bm{\hat{h}}}_{t,k}}$. With (\ref{eq:eq20})-(\ref{eq:eq22}), we can obtain the asymptotic value of the normalized achievable sum rate of the MMSE receiver according to the control convergence theorem \cite{6172680}, that is,
\begin{equation}\label{eq:eq56}
{{R}_{\text{MMSE},k}}-\frac{1}{N}\sum\limits_{t=1}^{{{N}_{\text{T}}}}{{{\log }_{2}}\left( 1+\text{Tr}\left({\bm{Q}_{t,k}}{\bm{T}_{t,k}} \right)  \right)}\xrightarrow{a.s.}0.
\end{equation}

\subsection{Asymptotic Achievable Sum Rate of the MRC-like Receiver}\label{sec:ss3_3}
To further improve the performance of the MRC receiver, we also consider an MRC-like receiver. To detect the $k$-th data symbol from the $t$-th transmit antenna, we first perform whitening of the interference-plus-noise with the 
statistics of the CSI before the MRC receiver. Rewrite (\ref{eq:eq10}) as
\begin{equation}\label{eq:eq26}
{{\bm{y}}_{k}}=\sqrt{{{E}_{\text{C}}}}{{\bm{\hat{h}}}_{t,k}}{{s}_{t,k}}+{{\bm{\breve{z}}}_{k}},
\end{equation}
where
\[{{\bm{\breve{z}}}_{k}} = \sqrt {{E_\text{C}}} \sum\limits_{l = 1,l \ne t}^{{{N_\text{T}}}} {{{{\bm{\hat h}}}_{l,k}}{s_{l,k}}}  + {{\bm{\hat z}}_k}.\]
The operation for performing whitening of the interference-plus-noise is denoted as
\begin{equation}\label{eq:eq27}
{{\bm{\bar{y}}}_{k}}={{\bm{R}}_{\text{W},t,k}}{{\bm{y}}_{k}},
\end{equation}
where
\begin{align*}
{{\bm{R}}_{{\text{W}},t,k}} \buildrel \Delta \over =& \left[ {{{\bm{{\rm I}}}_{{N_{\text{R}}}}} \otimes \left( {{E_\text{C}}\sum\limits_{{l_1} = 1,{l_1} \ne t}^{{{N_\text{T}}}} {{{{\bm{\hat R}}}_{{l_1},k}}} } \right.} \right.\\
&{\left. {\left. { + {E_\text{C}}\sum\limits_{{l_2} = 1}^{{{N_\text{T}}}} {{{{\bm{\tilde R}}}_{{l_2},k}}}  + {\sigma ^2}{{\bm{I}}_N}} \right)} \right]^{ - \frac{1}{2}}}.
\end{align*}
The signal model is then rewritten as:
\begin{equation}\label{eq:eq28}
{{\bm{\bar{y}}}_{k}}=\sqrt{{{E}_{\text{C}}}}{{\bm{\bar{H}}}_{k}}{{\bm{\bar{s}}}_{k}}+{{\bm{\bar{z}}}_{k}},
\end{equation}
where ${{\bm{\bar{H}}}_{k}}={{\bm{R}}_{\text{W},t,k}}{{\bm{\hat{H}}}_{k}}$. Then, the estimated data symbols are given by
\begin{equation}\label{eq:eq29}
{{\hat{s}}_{\text{MRC-like},t,k}}=\sqrt{{{E}_{\text{C}}}}\bm{\bar{h}}_{t,k}^{\text{H}}{{\bm{\bar{y}}}_{k}}.
\end{equation}
The SINR of the MRC-like receiver can be expressed as (\ref{eq:eq30}), which is shown on the bottom of next page.
\begin{figure*}[!b]
\hrulefill \normalsize
\begin{equation}\label{eq:eq30}
{{\gamma }_{\text{MRC-like},t,k}}=\frac{{{\left| \bm{\bar{h}}_{t,k}^{\text{H}}{{{\bm{\bar{h}}}}_{t,k}} \right|}^{2}}}{\bm{\bar{h}}_{t,k}^{\text{H}}\left( {{{\bm{\bar{H}}}}_{[t],k}}\bm{\bar{H}}_{[t],k}^{\text{H}}+{{\bm{R}}_{\text{W},t,k}}{{{\bm{\tilde{R}}}}_{k}}\bm{R}_{\text{W},t,k}^{\text{H}} \right){{{\bm{\bar{h}}}}_{t,k}}}.
\end{equation}
\end{figure*}
Similar to the MRC receiver (see Appendix \ref{sec:Appendix_I}), by using Theorems 3.4 and 3.7 of \cite{Couillet2011Random}, we have
\begin{equation}\label{eq:eq31}
{{\gamma }_{\text{MRC-like},t,k}}\xrightarrow{a.s.}{{\bar{\gamma }}_{\text{MRC-like},t,k}},
\end{equation}
where
\begin{equation}\label{eq:eq32}
\begin{aligned}
&{{\bar{\gamma }}_{\text{MRC-like},t,k}}={{N}_{\text{R}}}\text{Tr}\left( {{\left( {{{\bm{\hat{R}}}}_{t,k}} \right)}^{\frac{1}{2}}} \right. \\
&\left. \times {{\left( \sum\limits_{{{l}_{1}}=1,{{l}_{1}}\ne t}^{{{N}_{\text{T}}}}{{{{\bm{\hat{R}}}}_{{{l}_{1}},k}}}+\sum\limits_{{{l}_{2}}=1}^{{{N}_{\text{T}}}}{{{{\bm{\tilde{R}}}}_{{{l}_{2}},k}}}+\frac{1}{{{\gamma }_{\text{C}}}}{{\bm{I}}_{N}} \right)}^{\text{-}1}}{{\left( {{{\bm{\hat{R}}}}_{t,k}} \right)}^{\frac{1}{2}}} \right).
\end{aligned}
\end{equation}
Define
\begin{equation}\label{eq:eq33}
{{R}_{\text{MRC-like},k}}=\frac{1}{N}\sum\limits_{t=1}^{{{N}_{\text{T}}}}{{{\log }_{2}}\left( 1+{{\gamma }_{\text{MRC-like},t,k}} \right)}.
\end{equation}
Additionally, by the control convergence theorem,
\begin{equation}\label{eq:eq34}
{{R}_{\text{MRC-like},k}}-\frac{1}{N}\sum\limits_{t=1}^{{{N}_{\text{T}}}}{{{\log }_{2}}\left( 1+{{{\bar{\gamma }}}_{\text{MRC-like},t,k}} \right)}\xrightarrow{a.s.}0.
\end{equation}
We will show in the simulations that when $N$ is large, the performance of the 
MRC-like receiver will approach that of the MMSE receiver.

\section{Doppler Diversity Order and Coding Gain Loss of the MRC-like Receiver}\label{SEC:S4}
In this section, we first obtain the average SER expression when the system uses the MRC-like receiver. Then, based on the SER expression, we study the maximum Doppler diversity order and the minimum coding gain loss that can be achieved in this case.

\subsection{Average SER of MRC-like Detection}\label{sec:ss4_1}
For the convenience of analysis, we assume that the specific CSI corresponding to other transmit antennas is unknown when estimating the data symbols from the $t$-th antenna and that ${{\hat{s}}_{\text{MRC-like},t,k}}$ obeys a Gaussian distribution with ${{s}_{t,k}}$ and ${{\bm{\bar{h}}}_{t,k}}$ as conditions. The conditional mean and conditional variance of the decision variable can be given by
\[{{u}_{\hat{s}|\bm{\bar{h}},s,t,k}}={{E}_{\text{C}}}\bm{\bar{h}}_{t,k}^{\text{H}}{{\bm{\bar{h}}}_{t,k}}{{s}_{t,k}},\]
\[\sigma _{\hat{s}|\bm{\bar{h}},s,t,k}^{2}={{E}_{\text{C}}}\bm{\bar{h}}_{t,k}^{\text{H}}{{\bm{\bar{h}}}_{t,k}}.\]
\begin{theorem}\label{THEOREM 2}
Under the above conditions, the average symbol error rate of the system is
\begin{equation}\label{eq:eq35}
{{\bar{P}}_{\text{e},t,k}}=\frac{1}{\pi }{{\int_{0}^{\Theta }{\left[ \det \left( {{\bm{I}}_{N}}+\frac{{{C}_{M}}}{{{\sin }^{2}}\theta }{{\bm{A}}_{t,k}} \right) \right]}}^{-{{N}_{R}}}}d\theta,
\end{equation}
where $\Theta {=}\pi {-}\pi {/M}$,${{C}_{M}}={{\sin }^{2}}\left( \pi {/M} \right)$, and
\begin{align}\label{eq:eq36}
{{\bm{A}}_{t,k}}={{\bm{\hat{R}}}_{t,k}}{{\left( \sum\limits_{{{l}_{1}}=1,{{l}_{1}}\ne t}^{{{N}_{\text{T}}}}{{{{\bm{\hat{R}}}}_{{{l}_{1}},k}}}+\sum\limits_{{{l}_{2}}=1}^{{{N}_{\text{T}}}}{{{{\bm{\tilde{R}}}}_{{{l}_{2}},k}}}+\frac{1}{{{\gamma }_{\text{C}}}}{{\bm{I}}_{N}} \right)}^{\text{-}1}}.
\end{align}
\end{theorem}
\begin{IEEEproof}
See Appendix \ref{sec:Appendix_III}.
\end{IEEEproof}

\subsection{Doppler Diversity Order of MRC-like Detection}\label{sec:ss4_2}
Based on Jensen's inequality and the fact that ${{\sin }^{2}}\theta \le 1$, the upper and lower bounds of $\log \left( {{{\bar{P}}}_{\text{e},t,k}} \right)$ can be obtained.
\begin{equation}\label{eq:eq37}
{{\eta }_{\text{U}}}=\log v-{{N}_\text{R}}\log [\det ({{\bm{I}}_{N}}+{{C}_{{M}}}{{\bm{A}}_{t,k}})],
\end{equation}
\begin{equation}\label{eq:eq38}
{{\eta }_{\text{L}}}=\log v-\frac{{{N}_\text{R}}}{\pi v}\int_{0}^{\pi v}{\log [\det ({{\bm{I}}_{N}}+\frac{{{C}_{{M}}}}{{{\sin }^{2}}\theta }{{\bm{A}}_{t,k}})]}d\theta,
\end{equation}
where $v=1-1/M$. The proof is similar to Lemma 1 in \cite{Zhou2015High}. From Lemma 2 of \cite{Gazzah2001Asymptotic}, when $N{{N}_{\text{R}}}\to \infty $:
\begin{equation}\label{eq:eq39}
{{\bm{A}}_{t,k}}\xrightarrow{a.s.}\bm{U}_{N}^{\text{H}}{{\bm{D}}}{{\bm{U}}_{N}}.
\end{equation}
${{\bm{U}}_{N}}$ is an $N$-dimensional unitary discrete Fourier transform matrix, and ${{\bm{D}}}$ is a diagonal matrix with
\begin{align*}
&{({{\bm{D}}})_{n,n}}{\rm{ = }}\bigggl( {{{N_\text{T}}} - 1} \bigggr.\\
&{\rm{ + }}{\bigggl. {{{\left( {\frac{{{\Lambda _{{\text{PP}}}}^2(2\pi \frac{{n - 1}}{N})}}{{\frac{{{{N_\text{T}}}}}{{{\gamma _{\text{P}}}}}{\Lambda _{{\text{PP}}}}(2\pi \frac{{n - 1}}{N}) + \frac{1}{{{\gamma _{\text{C}}}}}{\Lambda _{{\text{PP}}}}(2\pi \frac{{n - 1}}{N}) + \frac{1}{{{\gamma _{\text{P}}}{\gamma _{\text{C}}}}}}}} \right)}^{{\rm{ - }}1}}} \bigggr)^{{\rm{ - 1}}}},
\end{align*}
\[{{\Lambda }_{\text{PP}}}(\Omega )=\frac{2\text{rect}(\frac{\Omega }{{4\pi {f_{\text{D}}}}{{T}_{\text{P}}}})}{\sqrt{{{({2\pi {f_{\text{D}}}}{{T}_{\text{P}}})}^{2}}-{{\Omega }^{2}}}},\;\;{-}\pi \le \Omega \le \pi.\]
It can be found that ${{\bm{D}}}$ is not related to $t$ and $k$, and the average SER at this time is ${{\bar{P}}_{\text{e}}}$. Define the energy required to transmit a coded data symbol as
$${{E}_{0}} = \left( {1/K} \right){{E}_{\text{P}}}\text{+}{{E}_{\text{C}}},$$
and the corresponding SNR as ${{\gamma }_{0}}={{E}_{0}}/{{\sigma }^{2}}$.
Define
$${\Psi _{t,k}}({\gamma _0}) = \det ({{\bm{I}}_N} + c{{\bm{A}}_{t,k}}),$$
where $c$ is a constant. Let
$$\delta ={2\pi {f_{\text{D}}}}{{T}_{\text{P}}}, \;\; {{\gamma }_{\text{t}}}={{N}_{\text{T}}}{{\gamma }_{\text{C}}}+{{\gamma }_{\text{P}}}, \;\; {{\gamma }_{\text{P}}}=b{{\left( {{\gamma }_{\text{C}}} \right)}^{\xi }},$$
where $b$ and $\xi $ are constants. According to Appendix D of \cite{Zhou2015High}, it can be inferred that:
\begin{align}\label{eq:eq40}
&\underset{N\to \infty }{\mathop{\lim }}\,\frac{\log \left( \Psi_{t,k}\left( {{\gamma }_{0}} \right) \right)}{N} \nonumber \\
&=\frac{\delta }{\pi }\log \left( 4{{\gamma }_{\text{P}}}{{\gamma }_{\text{C}}}c \right)-\frac{\delta \log \left( 2\delta  \right)-2\delta }{\pi }\nonumber \\
&\quad-\frac{\delta \log \left( 2{{\gamma }_{\text{t}}} \right)}{\pi }-\left( {{\gamma }_{\text{t}}}-\frac{\sqrt{4\gamma _{\text{t}}^{2}-{{\delta }^{2}}}}{2} \right)\nonumber \\
&\quad-\frac{\sqrt{4\gamma _{\text{t}}^{2}-{{\delta }^{2}}}}{\pi }\arctan \left( \sqrt{\frac{{{\delta }^{2}}}{4\gamma _{\text{t}}^{2}-{{\delta }^{2}}}} \right).
\end{align}
Consider the definition of the normalized Doppler diversity order in \cite{Zhou2015High}:
\begin{equation}\label{eq:eq41}
D=-\underset{\begin{smallmatrix}
 {{\gamma }_{0}}\to \infty  \\
 N\to \infty
\end{smallmatrix}}{\mathop{\lim }}\,\frac{\log \left( {{{\bar{P}}}_{\text{e}}} \right)}{N{{T}_{\text{P}}}\log \left( {{\gamma }_{0}} \right)}.
\end{equation}
(\ref{eq:eq37}) and (\ref{eq:eq38}) are substituted into (\ref{eq:eq41}) to calculate the upper and lower bounds of the order of the Doppler diversity ${{D}_{\text{U}}}$ and ${{D}_{\text{L}}}$. Let $c$ be ${{C}_{{M}}}$ in the upper bound and ${{C}_{{M}}}/{{\sin }^{2}}\theta$ in the lower bound, and combine the results with formulas (\ref{eq:eq39}) and (\ref{eq:eq40}). As in \cite{Mahamadu2018Fundamental}, we obtain
\begin{align}\label{eq:eq42}
{{D}_{\text{U}}}={{D}_{\text{L}}}&=-\underset{\begin{smallmatrix}
 {{\gamma }_{0}}\to \infty  \\
 N\to \infty
\end{smallmatrix}}{\mathop{\lim }}\,\frac{{{N}_{\text{R}}}\log \left( \Psi_{t,k} \left( {{\gamma }_{0}} \right) \right)}{N{{T}_{\text{P}}}\log \left( {{\gamma }_{0}} \right)}\nonumber \\
&=\left\{ \begin{matrix}
   2{{f}_{\text{D}}}{{N}_{\text{R}}}\xi ,\xi \le 1  \\
   2{{f}_{\text{D}}}{{N}_{\text{R}}}\frac{1}{\xi },\xi >1  \\
\end{matrix} \right..
\end{align}
Then, $D$ can be obtained by the clamping theorem. Observing (\ref{eq:eq42}), we know that when $\xi =1$, the maximum normalized Doppler order is reached, and the maximum value is $2{{f}_{\text{D}}}{{N}_{\text{R}}}$.

\subsection{Coding Gain Loss in MRC-like Detection}\label{sec:ss4_3}
In this subsection, the coding gain loss caused by imperfect CSI in MRC-like detection is derived, and the corresponding energy allocation scheme of minimum coding gain loss is obtained. As seen from \cite{Mahamadu2018Fundamental}, the coding gain in logarithmic form is:
\begin{equation}\label{eq:eq43}
\log C=-\underset{\begin{smallmatrix}
 {{\gamma }_{0}}\to \infty  \\
 N\to \infty
\end{smallmatrix}}{\mathop{\lim }}\,\left( \frac{\log \left( {{{\bar{P}}}_{\text{e}}} \right)}{N{{T}_{\text{P}}}D}+\log {{\gamma }_{0}} \right).
\end{equation}
Let $\xi =1$, combine (\ref{eq:eq37}) and (\ref{eq:eq38}) and simplify to obtain the lower and upper bounds in the presence of imperfect CSI:
\begin{align}\label{eq:eq44}
\log {{C}_{\text{L}}}=&\log 2{{C}_{{M}}}+\log \left( \frac{Kb}{\left( K+b \right)\left( {{N}_{\text{T}}}+b \right)} \right)+1\nonumber \\
&-\log 2\delta,
\end{align}
\begin{equation}\label{eq:eq45}
\log {{C}_{\text{U}}}=\log {{C}_{\text{L}}}-\frac{1}{2{{f}_{\text{D}}}{{T}_{\text{P}}}\pi v}\int_{0}^{\pi v}{2\log \left( \sin \theta  \right)d\theta }.
\end{equation}
In the case of perfect CSI, channel estimation is not required, ${{\gamma }_{0}}={{\gamma }_{\text{C}}}$,
\[{{({{\bm{D}}})}_{n,n}}={{\left( {{N}_{\text{T}}}-1+{{\Lambda }_{\text{PP}}^{-1}}(2\pi \frac{n-1}{N}) \right)}^{{-}1}},\]
corresponding to
\begin{equation}\label{eq:eq46}
\underset{N\to \infty }{\mathop{\lim }}\,\frac{\log \left( \Psi '_{t,k}\left( {{\gamma }_{0}} \right) \right)}{N}=\frac{\delta }{\pi }\log \left( 2c{{\gamma }_{\text{C}}} \right)-\frac{\delta \log 2\delta -\delta }{\pi }.
\end{equation}
The upper and lower bounds in the presence of perfect CSI are:
\begin{equation}\label{eq:eq47}
\log {{C}_{\text{L}}'}=\log \left( 2{{C}_{{M}}} \right)\text{+}1-\log 2\delta,
\end{equation}
\begin{equation}\label{eq:eq48}
\log {{C}_{\text{U}}'}=\log {{C}_{\text{L}}'}-\frac{1}{2{{f}_{\text{D}}}{{T}_{\text{p}}}\pi v}\int_{0}^{\pi v}{2\log \left( \sin \theta  \right)d\theta }.
\end{equation}
According to the definition of \cite{Mahamadu2018Fundamental},
\begin{align}\label{eq:eq49}
 {{\Upsilon }_{\text{Loss}}}\text{(dB)} &= 10{{\log }_{10}}{{C}_{\text{U}}'}\text{-10}{{\log }_{10}}{{C}_{\text{U}}}\nonumber\\
 &= \text{10}{{\log }_{10}}{{C}_{\text{L}}'}\text{-10}{{\log }_{10}}{{C}_{\text{L}}}\nonumber\\
 &= 10{{\log }_{10}}\left[ \frac{(K+b)({{N}_{\text{T}}}+b)}{Kb} \right].
\end{align}
The contents in the brackets are ${{N}_{\text{T}}}/b+b/K+\left( {{N}_{\text{T}}}+K \right)/K$. It can be seen from the definition that all values are greater than 0. When ${{N}_{\text{T}}}/b=b/K$, that is, $b=\sqrt{{{N}_{\text{T}}}K}$, the minimum value of (\ref{eq:eq49}) is achieved and the minimum value is $\text{2}0{{\log }_{10}}\left( 1+\sqrt{{{N}_{\text{T}}}/K} \right)$dB. Combining the above results, the maximum normalized Doppler diversity order is determined by the maximum Doppler spread ${{f}_{\text{D}}}$ and the number of receive antennas ${{N}_{\text{R}}}$. The minimum coding gain loss is determined by the number of transmit antennas ${{N}_{\text{T}}}$ and the number of data symbols in a block $K$.

\section{Numerical Results}\label{sec:s5}
In this section, numerical results are illustrated to investigate the performance of the MIMO system with Doppler diversity. Unless otherwise stated, 
the system transmission rate is ${{10}^{5}}$ Sym/s, the system operates at 1.9 GHz, ${{N}_{\text{T}}}=4$, ${{N}_{\text{R}}}=8$, $K=16$, $N\text{=15}$, and the modulation type is 4PSK. The pilot SNR ${{\gamma }_{\text{P}}}$ and data SNR ${{\gamma }_{\text{C}}}$ are both 10 dB.

\subsection{Performance of the Repetition Code in High-Speed Scenarios}\label{sec:ss5_1}

To study the performance of the repetition code, we compare the SER curves of the repetition code and the well-known Alamouti space-time code in the presence of perfect CSI. Both the transmitter and receiver ends use two antennas. It can be seen from Fig.\ref{fig2} that when the Doppler spread is small, the SER of the Alamouti code is lower than that of the repetition code with ${N}=2$. However, when the number of repetition times increases to 4, the SER of the repetition coding decreases significantly and is better than that of the Alamouti code. When the Doppler spread becomes larger, the SER of the Alamouti code becomes worse because of the destruction of spatial orthogonality, while the SER of the repetition coding improves because it can obtain higher Doppler diversity. It can also be found from Fig.\ref{fig2} that for a higher repetition time, more performance gain can be obtained from a higher Doppler spread. Actually, the repetition code can also be combined with the Alamouti code to exploit both Doppler diversity and spatial diversity.

\Figure[!h][width=80mm]{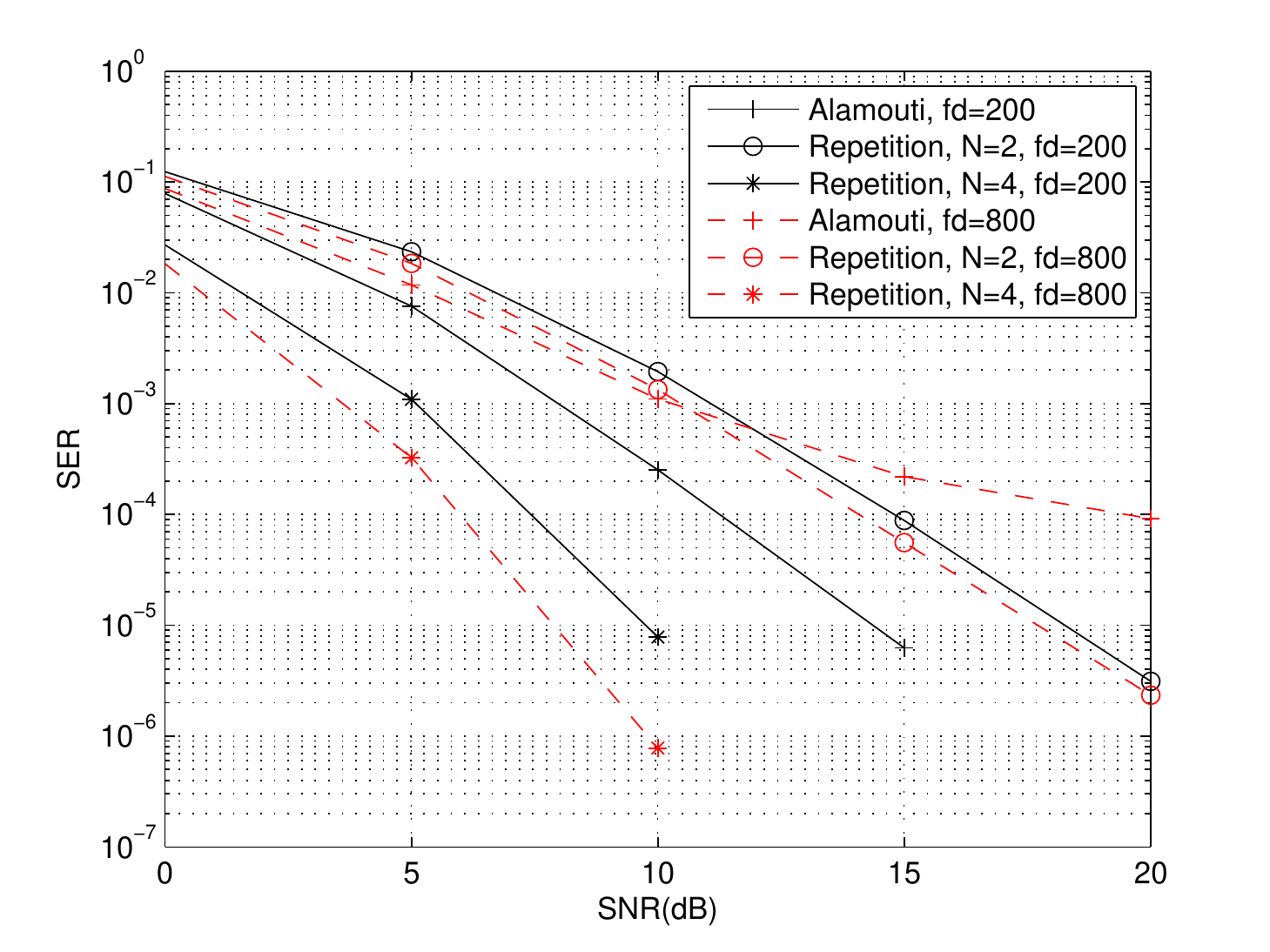}
{SER performance of the repetition code in the presence of perfect CSI\label{fig2}}

\subsection{Asymptotic Analysis Results of the Normalized Achievable Sum Rates}\label{sec:ss5_2}

\Figure[!h][width=80mm]{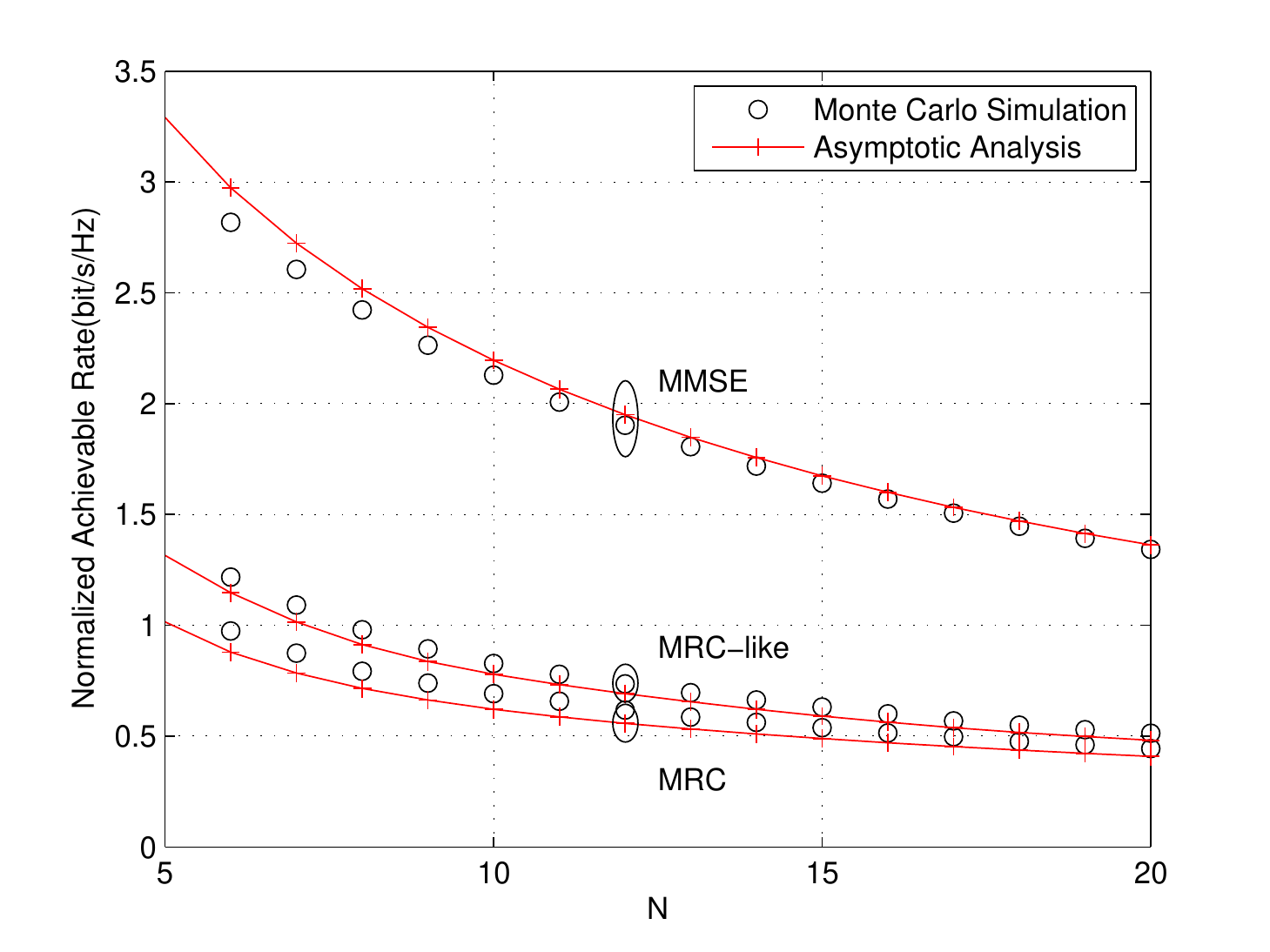}
{Normalized achievable sum rates as functions of $N$\label{fig3}}

Without loss of generality, we analyze the ergodic normalized achievable sum rates and corresponding asymptotic analysis results of the three detection algorithms for different repetition times $N$. The Doppler spread is set to 200 Hz, corresponding to a speed of 113.6 km/h. The number of receive antennas is ${{N}_{\text{R}}}\text{=4}$. It can be seen from Fig.\ref{fig3} that the MMSE receiver has the best performance. As $N$ increases, the normalized sum rate of each receiver decreases. This result means that the spectral efficiency of the system will decrease with an increase in the number of repetitions. By comparing the achievable sum rates of the three receivers, it is found that the performance gap narrows gradually with an increase in the repetition time. On the other hand, it can also be seen that the ergodic results are in good agreement with the corresponding analytical asymptotic results, which proves the correctness of the derivations in Section \ref{sec:s3}.

\subsection{Influence of the Parameters on the Normalized Achievable Sum Rates}\label{sec:ss5_3}

In this subsection, the influence of various parameters on the spectral efficiency in the presence of imperfect CSI is analyzed based on the numerical results.

\Figure[!h][width=80mm]{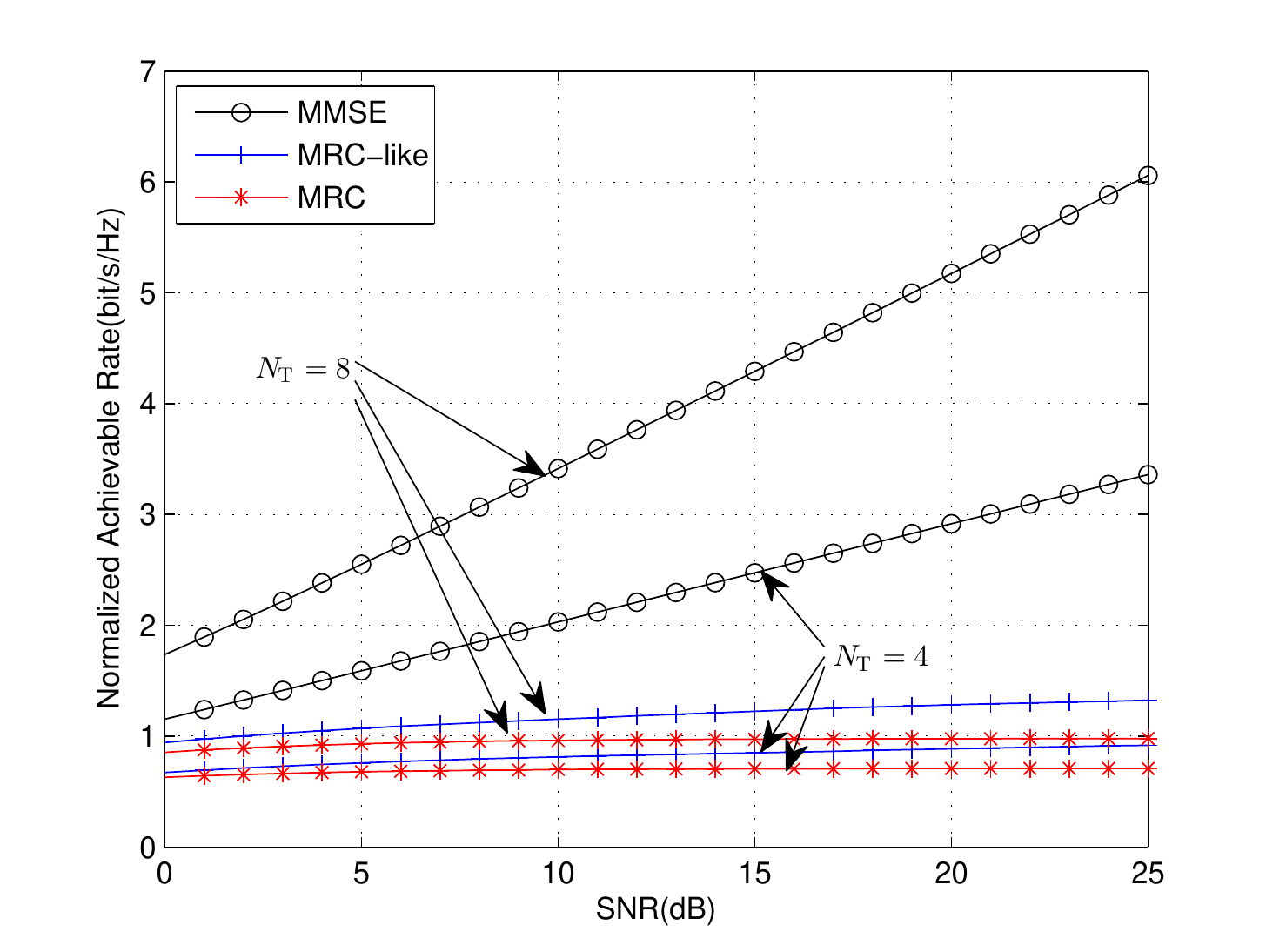}
{Normalized achievable sum rates as functions of the SNR at low speed\label{fig4}}

Fig.\ref{fig4} shows the change in the normalized sum rates with the SNR at low mobile speed. Since the correctness of the analytical results is verified above, the analytical results are directly used here for the calculation. The Doppler spread of the system is 200 Hz. The analytical results of the three receivers with different SNRs when ${{N}_{\text{T}}}=4$ and ${{N}_{\text{T}}}=8$ are calculated. Selecting the number of receive antennas as ${{N}_{\text{R}}}\text{=8}$ ensures the normal operation of the space multiplexing receiver. As can be seen from the figure, when each transmit antenna transmits different code words, the system can obtain greater spatial multiplexing gain and a higher achievable sum rate when the number of transmit antennas increases. By comparing the three receivers, we can also find that the normalized sum rate of the MMSE receiver increases with an increase in the pilot and data symbol SNR for any ${{N}_{\text{T}}}$, while the achievable rate of the MRC receiver hardly changes when the SNR increases to a certain extent; this result is due to the poor ability of the MRC receiver to suppress interference, and the interference will limit the performance of the system when $N{{N}_{\text{R}}}$ is limited and fixed. Considering the interference from other antennas, the sum rate of the MRC-like receiver can keep increasing with 
an increase in the SNR, and the performance is between the performances of the other two receivers.

\Figure[!h][width=80mm]{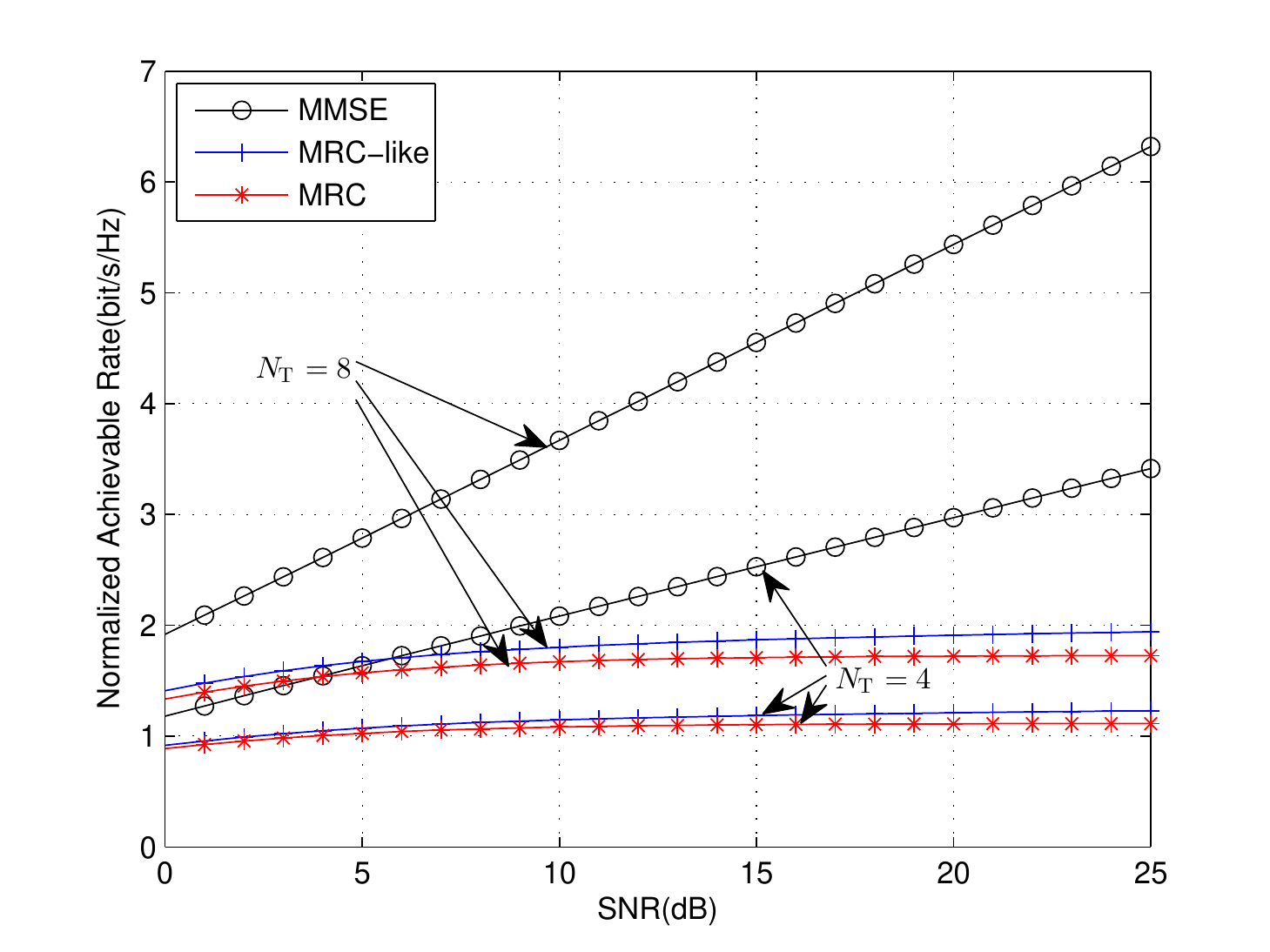}
{Normalized achievable sum rates as functions of the SNR at high speed\label{fig5}}

To study the system at high speed, we keep the other parameters unchanged in Fig.\ref{fig5} and increase the Doppler spread to 1000 Hz to obtain the achievable sum rates of the three receivers in high-mobility scenarios. As can be seen from the figure, since the pilot interval still satisfies the requirement of $\left( {{N}_{\text{T}}}+K \right)T\le {0.5}/{{{f}_\text{D}}}\;$, the system can still operate normally. Similar to the low speed case, the achievable rates of the three receivers increase with the number of transmit antennas. Moreover, similar to the low speed case, the achievable sum rates of the three receivers increase with the number of transmit antennas, the performance of the MMSE receiver steadily increases with an increase in the SNR, the performance of the MRC receiver is limited by the interference, and 
the performance of the MRC-like receiver is between the performances of the other two receivers. By comparing Fig.\ref{fig4} and Fig.\ref{fig5}, we find that MIMO systems with repetition coding can achieve a higher rate in high-mobility scenarios, because at the same sampling rate, the correlation of the channel information corresponding to the repetitive coded data is weakened.

\Figure[!h][width=80mm]{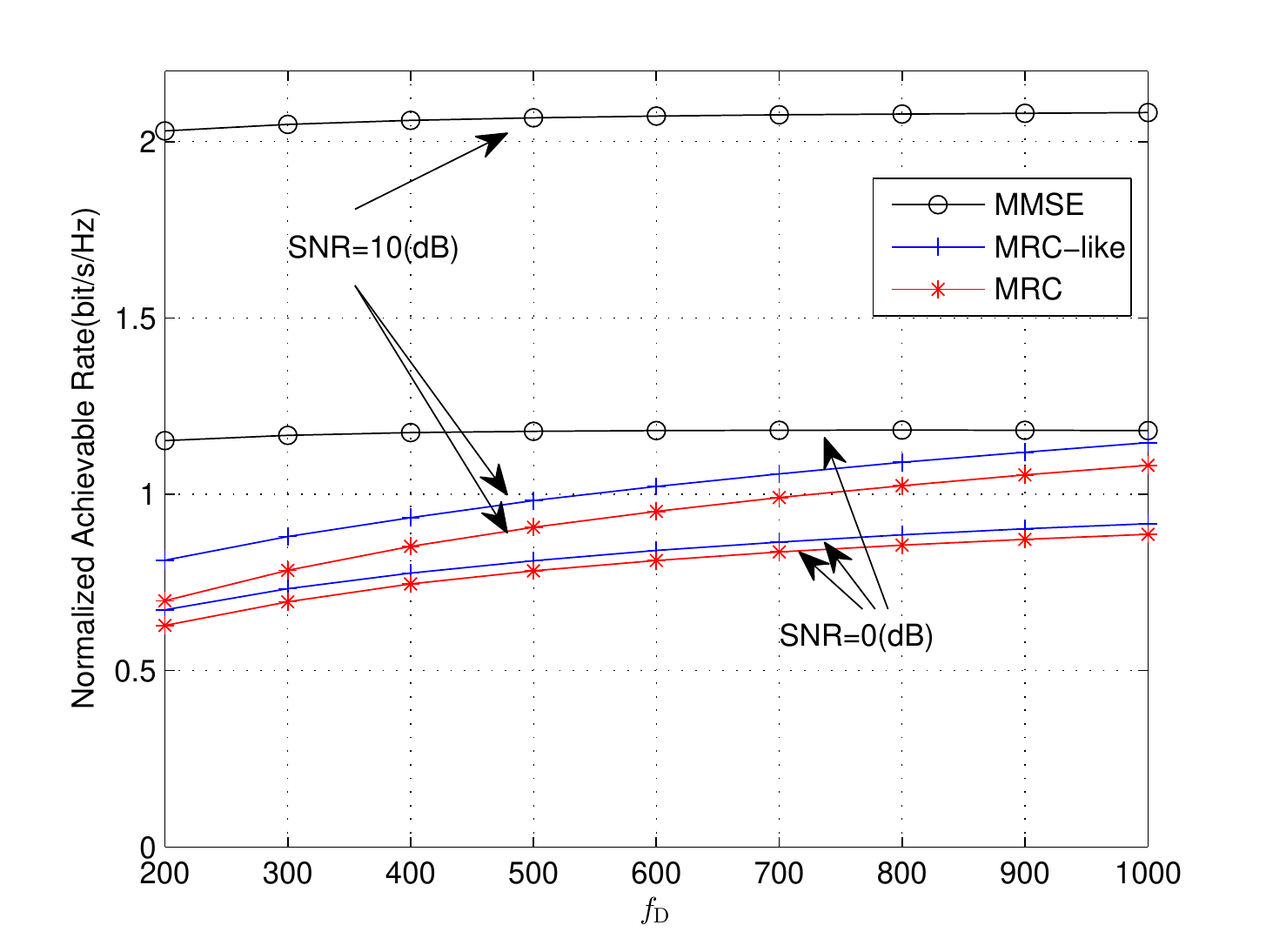}
{Normalized achievable sum rates as functions of ${{f}_\text{D}}$\label{fig6}}

To obtain a more intuitive understanding of the influence of different ${{f}_\text{D}}$ on the normalized achievable sum rate, we calculate the normalized achievable sum rates for different ${{f}_\text{D}}$, as shown in Fig.\ref{fig6}. The figure shows that in the system described in this paper, an increase in ${{f}_\text{D}}$ can significantly improve the performance of the MRC receiver and 
MRC-like receiver but has little effect on the performance of the MMSE receiver.

\Figure[!h][width=80mm]{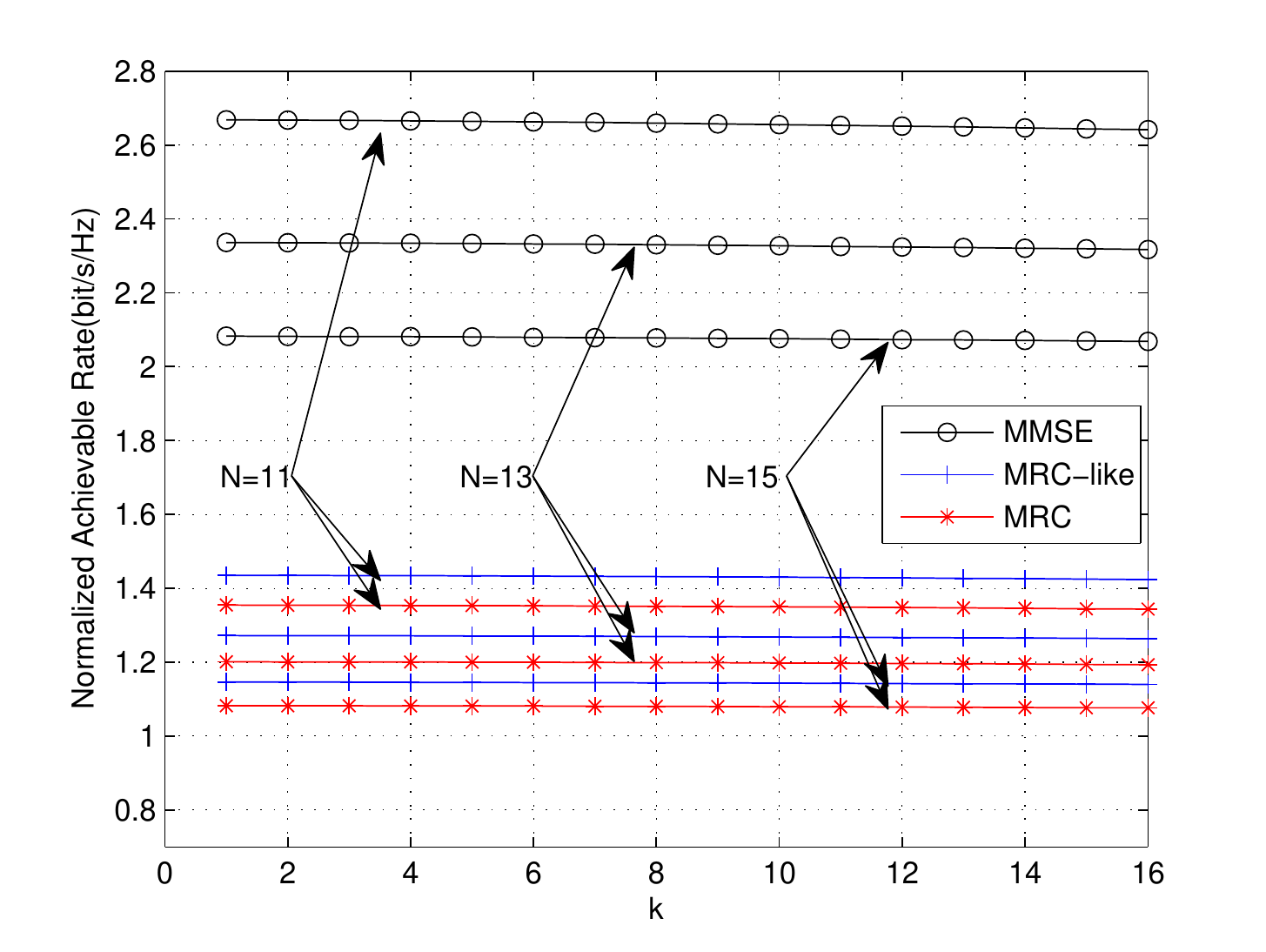}
{Normalized achievable sum rates as functions of $k$\label{fig7}}

In Fig.\ref{fig7}, we calculate the analytical results of the three receivers corresponding to different values of $k$ and $N$ in the high-mobility scene. From the figure, we can see that under the simulation environment that is set up, when the other conditions are fixed, the normalized achievable sum rates for different values of $k$ are basically the same.

\Figure[!h][width=80mm]{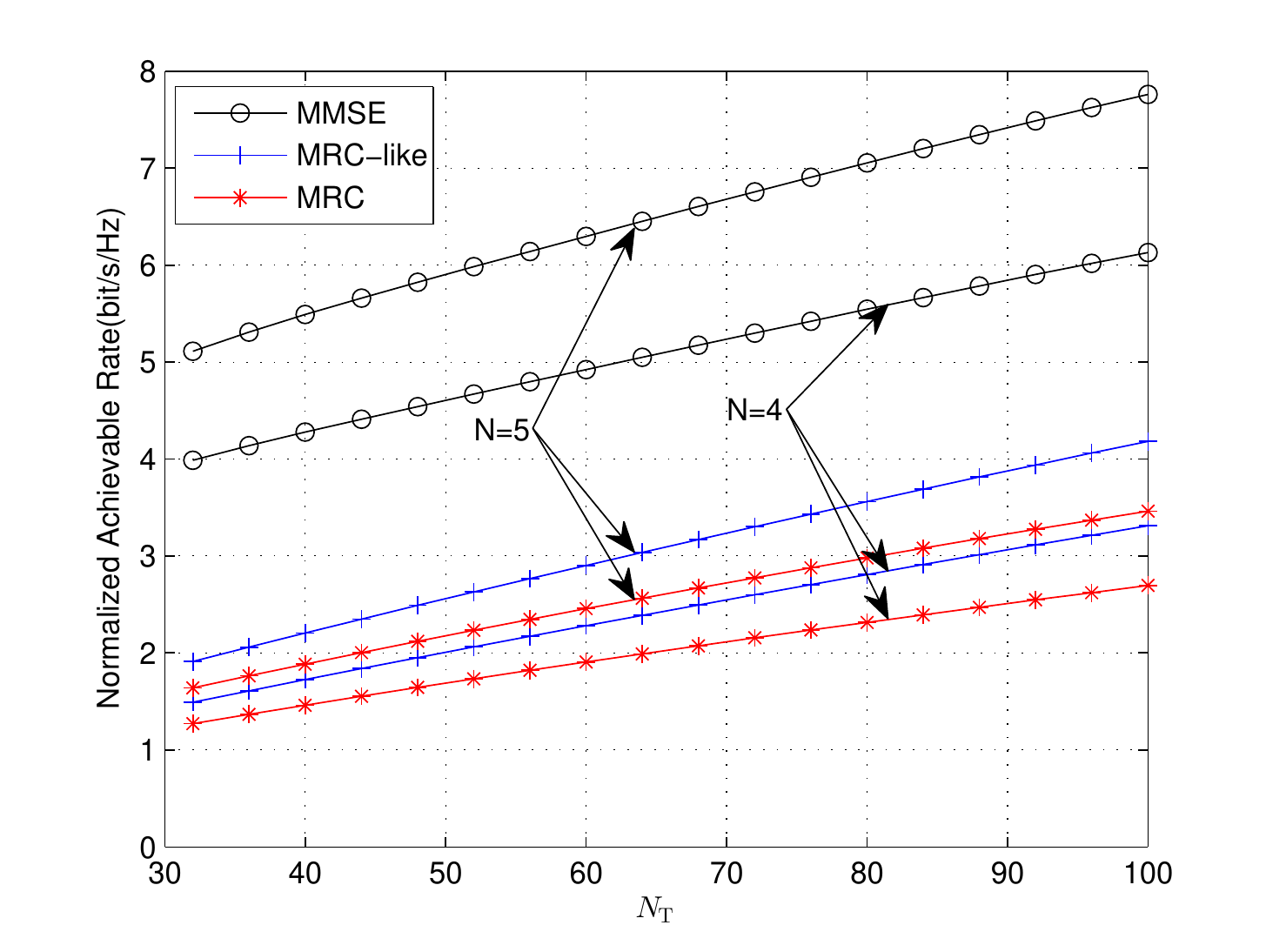}
{Normalized achievable sum rates as functions of ${{N}_{\text{T}}}$ at high speed\label{fig8}}

In Fig.\ref{fig8}, we calculate the normalized achievable sum rates of the three algorithms at high speed. To ensure normal operation of the system, we set the symbol transmission rate to $3\times {{10}^{5}}$ Sym/s and let $N{{N}_{\text{R}}}=100$. As can be seen from the figure, more spatial degrees of freedom from the increased number of transmit antennas will help to improve the overall performance of the system.

\Figure[!h][width=80mm]{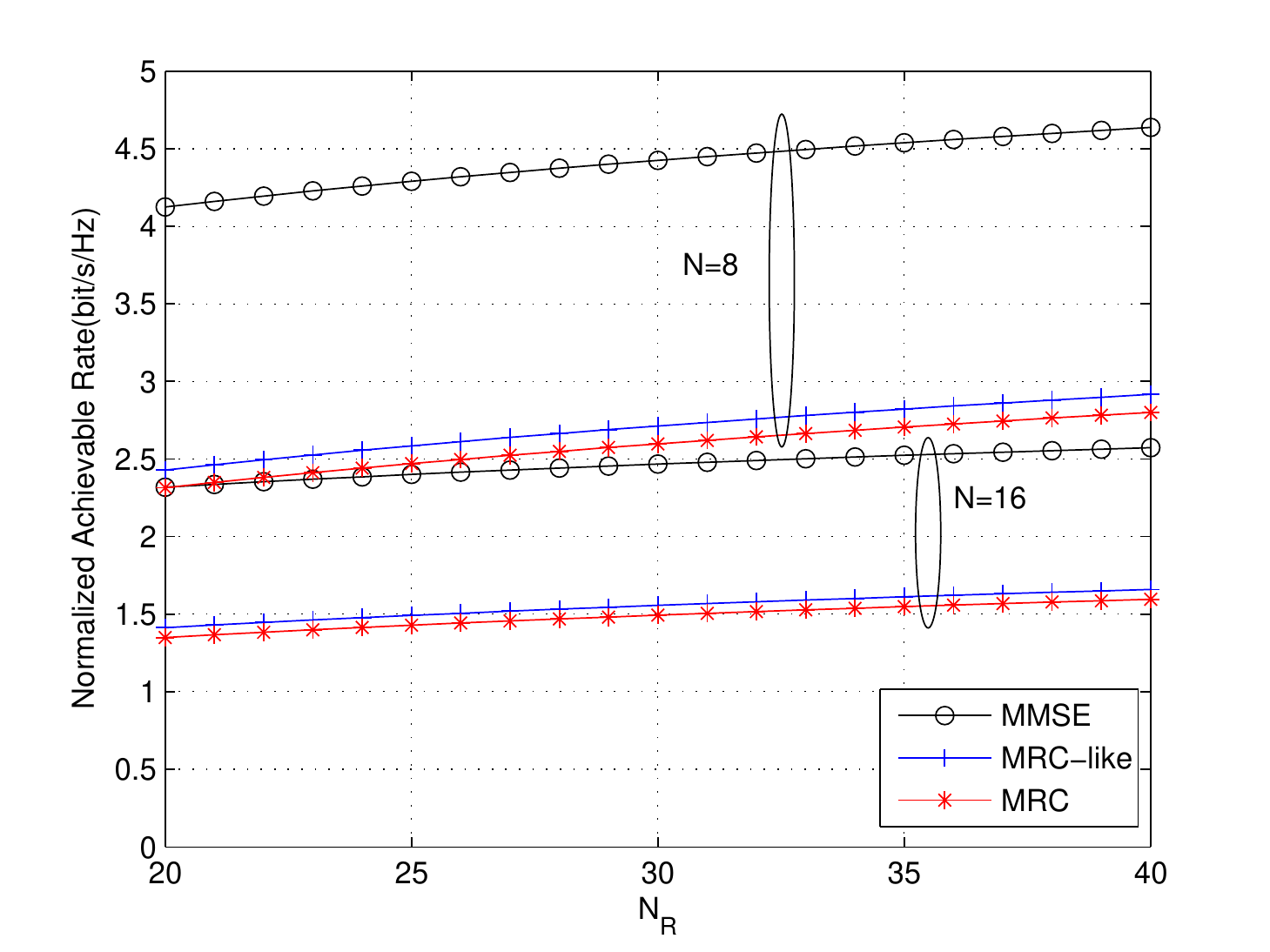}
{Normalized achievable sum rates as functions of ${{N}_{\text{R}}}$ at high speed\label{fig9}}

Fig.\ref{fig9} studies the normalized achievable sum rates at high speed and for large ${{N}_{\text{R}}}$. In this case, with an increase in the number of receive antennas, the system obtains more spatial degrees of freedom, and the system performance is improved. When the other conditions are fixed, the performance gap between the different algorithms is basically fixed.

\Figure[!h][width=80mm]{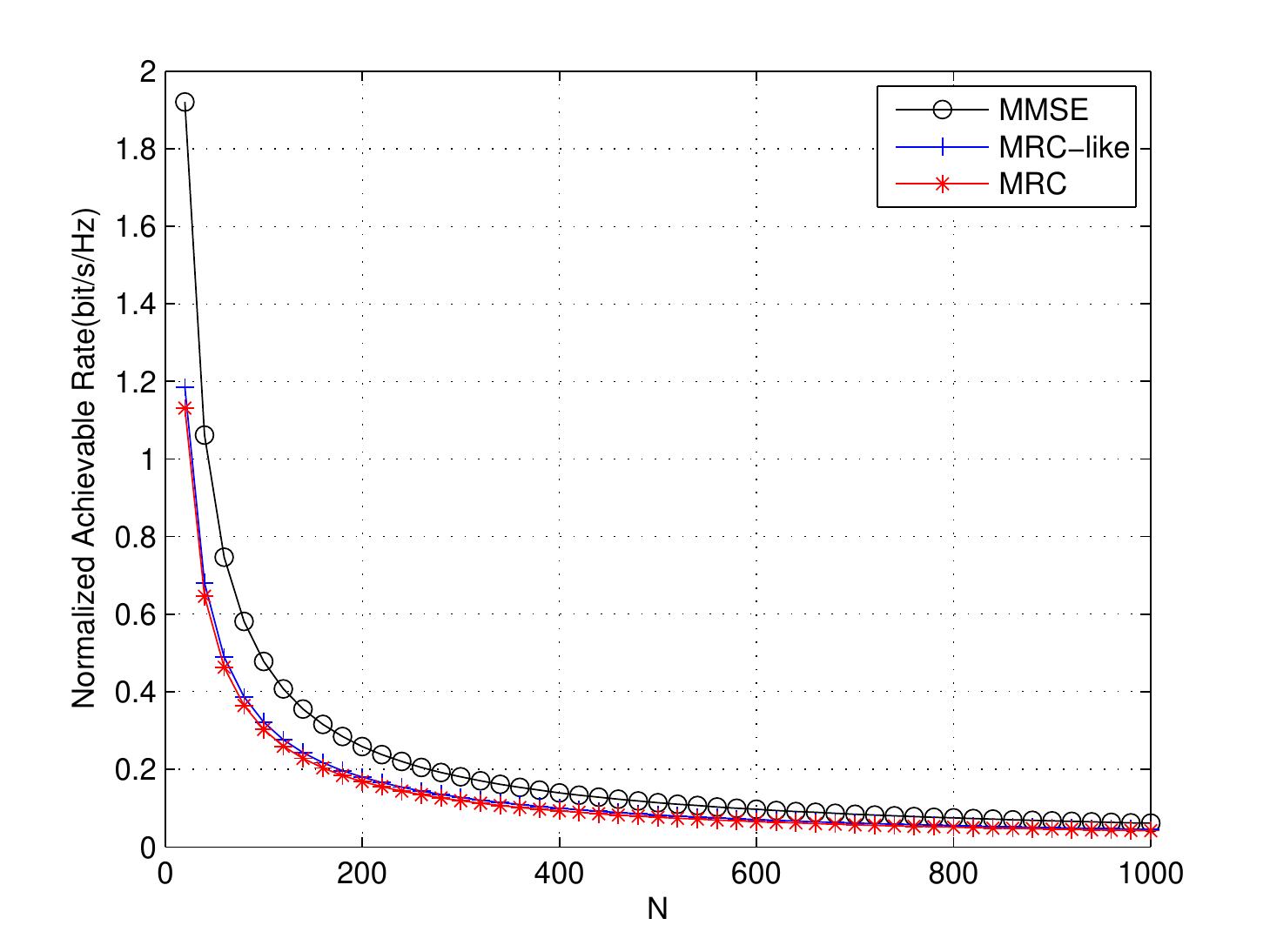}
{Normalized achievable sum rates as functions of $N$ at high speed\label{fig10}}

In Fig.\ref{fig10}, the normalized achievable sum rates at high speed are studied as a function of $N$, and ${{N}_{\text{R}}}$ is set to 20. The figure shows that the normalized achievable sum rates of different algorithms converge in the process of increasing $N$, which means that the performance gap of the three algorithms tends to disappear when $N$ tends to infinity.

\subsection{Asymptotic Analysis Results in Section \ref{SEC:S4}}\label{sec:ss5_4}

In this subsection, numerical results based on the analysis in section \ref{SEC:S4} are illustrated.

\Figure[!h][width=80mm]{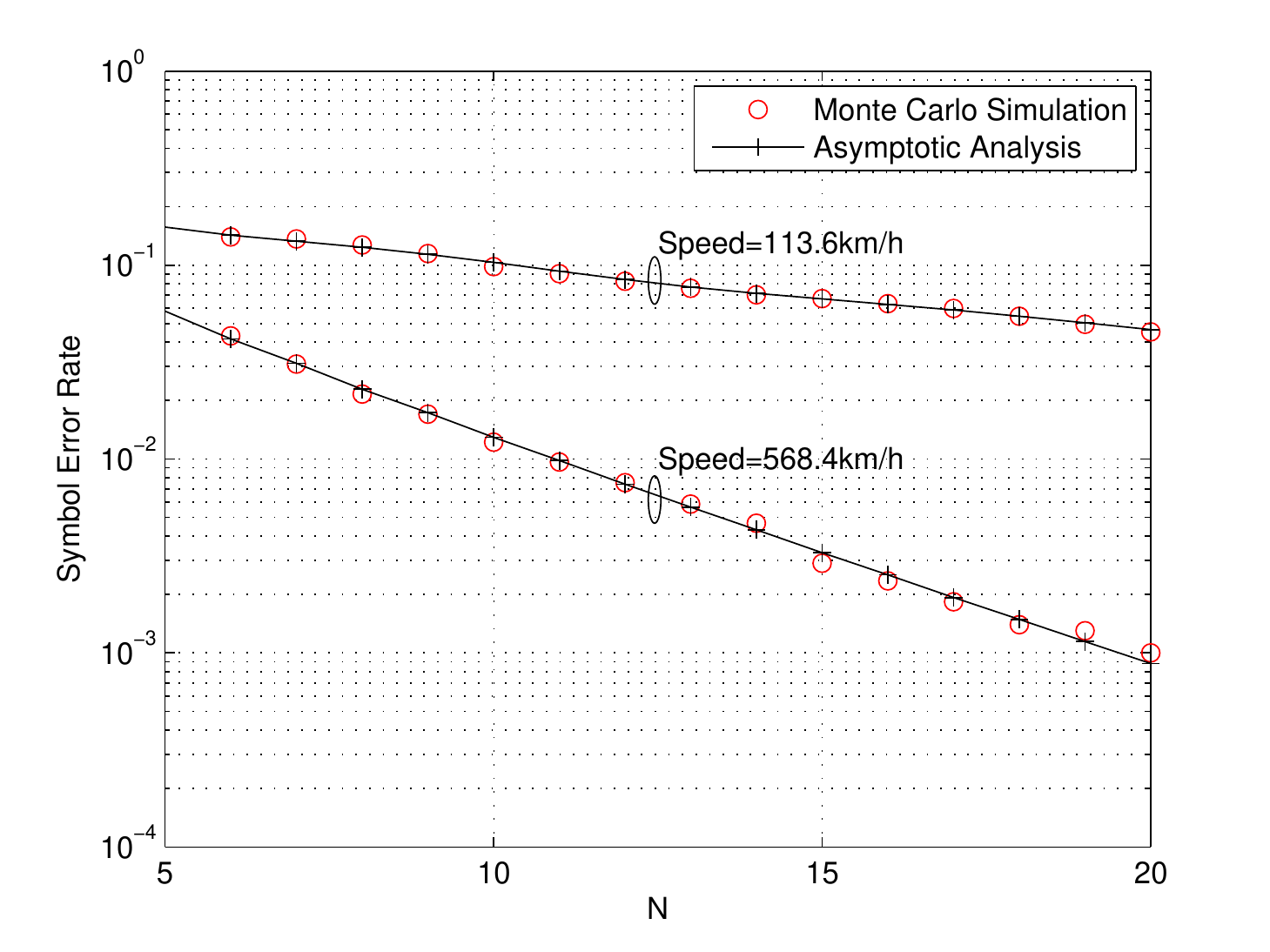}
{SER of the MRC-like receiver as functions of $N$\label{fig11}}

The average SER of the MRC-like receiver is calculated in Fig.\ref{fig11} by the Monte Carlo method and the analysis result (\ref{eq:eq35}). Set ${{N}_{\text{R}}}\text{=4}$. The analytical results match the simulation results well. In addition, as $N$ increases, the SER decreases steadily, and the speed of the decrease at high speed is faster, because the Doppler order that can be obtained at high speed is higher.

\Figure[!h][width=80mm]{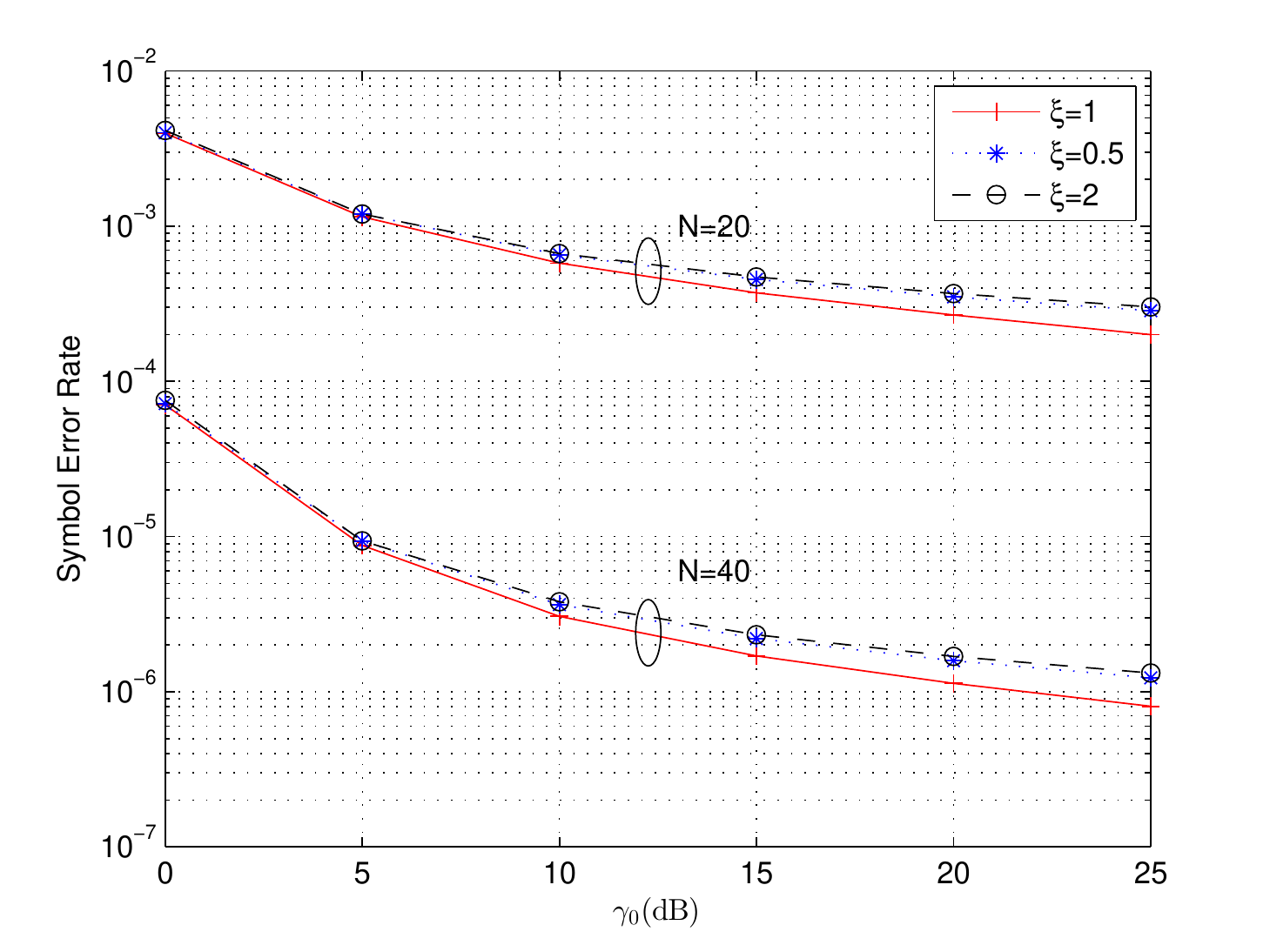}
{SER of the MRC-like receiver as functions of ${{\gamma }_{0}}$\label{fig12}}

Fig.\ref{fig12} shows the case where the SER changes with ${{\gamma }_{0}}$ at high speed when $N$ is finite. Set ${{N}_{\text{R}}}\text{=4}$ and $b=\sqrt{{{N}_{\text{T}}}K}$. The decreasing speed of the SER curve decreases with an increase in ${{\gamma }_{0}}$ but increases with an increase in $N$. The simulation results for different $\xi$ show that better performance can be achieved with $\xi=1$, which is consistent with the conclusion achieved for the case when $N$ tends to infinity.

\Figure[!h][width=80mm]{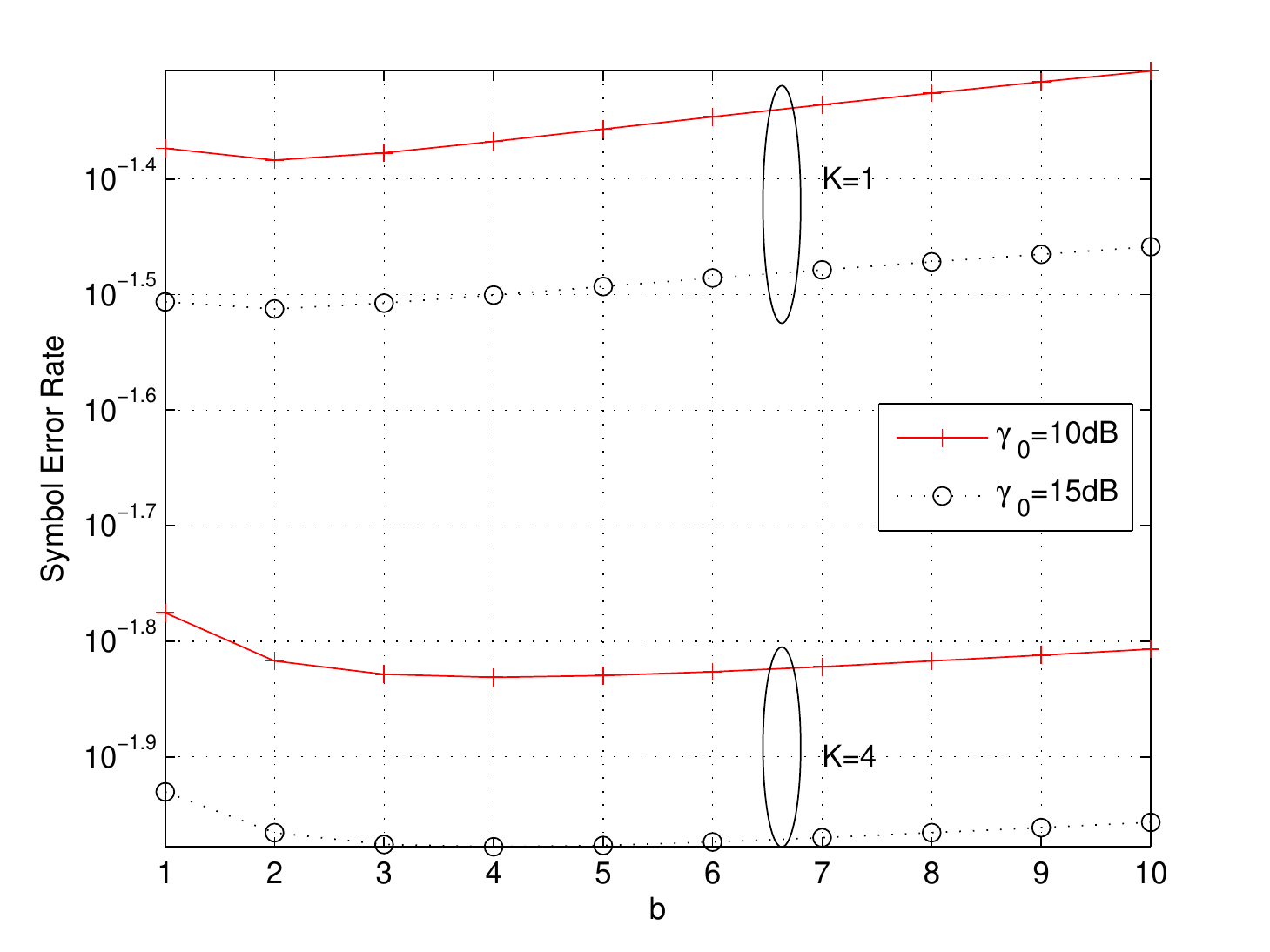}
{SER of the MRC-like receiver as functions of $b$\label{fig13}}

Fig.\ref{fig13} shows the SER as a function of $b$ at high speed in the case of a finite $N$. Set ${{N}_{\text{R}}}\text{=4}$ and $\xi=1$. Since there is no need to consider the pilot and data energy allocation in the presence of perfect CSI conditions, the SER performance is fixed, so the best performance in the presence of imperfect CSI corresponds to the minimum performance loss. The figure shows that when $N$ is finite, the system has the best performance when $b=\sqrt{{{N}_{\text{T}}}K}$, which is also consistent with the conclusion for the case when $N$ tends to infinity.

\section{Conclusions}\label{sec:s6}
In this paper, the performance of an MIMO system using repetition coding to achieve Doppler diversity in high-mobility scenarios is analyzed. Based on the characteristics of MMSE channel estimation, the equivalent system model is derived, and asymptotic expressions of the normalized achievable sum rates for MRC, MMSE and MRC-like receivers are obtained. Then, the expression of the average SER of the MRC-like receiver is derived, and the maximum normalized Doppler diversity order, the minimum coding gain loss and the corresponding conditions are obtained.

The effects of different system parameters on the system performance are studied. Increasing the number of transmit and receive antennas can provide more spatial degrees of freedom, and the system performance will be improved. In a high-mobility MIMO system using repetition coding, the MMSE receiver has a higher normalized achievable sum rate than the other two receivers when $N$ is finite. The gap between the different receivers tends to disappear when $N$ tends to infinity.
When the Doppler spread increases, the average SER of the MRC-like receiver will decrease. The larger the Doppler spread, the faster the average SER curve decreases with $N$, due to the higher Doppler diversity order.

\appendices
\section{Proofs of Theorem \ref{THEOREM 1}}\label{sec:Appendix_I}
Considering the numerator and denominator part of (\ref{eq:eq14}), we rewrite ${\bm{\hat h}}_{t,k}^{\text{H}}{{\bm{\hat h}}_{t,k}}$ as
\begin{align*}
&{\bm{\hat h}}_{t,k}^{\text{H}}{{\bm{\hat h}}_{t,k}}=\\
&N{{N_\text{R}}}{\left( {\frac{1}{{\sqrt {N{{N_\text{R}}}} }}{{{\bm{\mathord{\buildrel{\lower3pt\hbox{$\scriptscriptstyle\frown$}}
\over h} }}}_{t,k}}} \right)^{\text{H}}}\left( {{{\bm{I}}_{{{N_\text{R}}}}} \otimes {{{\bm{\hat R}}}_{t,k}}} \right)\left( {\frac{1}{{\sqrt {N{{N_\text{R}}}} }}{{{\bm{\mathord{\buildrel{\lower3pt\hbox{$\scriptscriptstyle\frown$}}
\over h} }}}_{t,k}}} \right).
\end{align*}
The elements in the vector $\frac{1}{{\sqrt {N{{N_\text{R}}}} }}{{\bf{\mathord{\buildrel{\lower3pt\hbox{$\scriptscriptstyle\frown$}}\over h} }}_{t,k}}$ are independently and identically distributed, the mean value is 0 and the variance is $\frac{1}{N{{N}_\text{R}}}$.
From Theorem 3.4 of \cite{Couillet2011Random}, when $N{{N}_\text{R}}\to \infty $,
$$\bm{\hat{h}}_{t,k}^{\text{H}}{{\bm{\hat{h}}}_{t,k}}\xrightarrow{a.s.}N{{N}_{\text{R}}}\left( \frac{1}{N{{N}_{\text{R}}}}\text{Tr}\left( {{\bm{I}}_{{{N}_{\text{R}}}}}\otimes {{{\bm{\hat{R}}}}_{t,k}} \right) \right).$$
Considering the characteristics of the Kronecker product with the identity matrix, the expression can be simplified to
\begin{equation}\label{eq:eq50}
\bm{\hat{h}}_{t,k}^{\text{H}}{{\bm{\hat{h}}}_{t,k}}\xrightarrow{a.s.}{{N}_{\text{R}}}\text{Tr}\left( {{{\bm{\hat{R}}}}_{t,k}} \right).
\end{equation}
For the denominator, we first rewrite the expression as
\begin{align*}
&{\bm{\hat{h}}_{t,k}^{\text{H}}\left( {{{\bm{\hat{H}}}}_{\left[ t \right],k}}\bm{\hat{H}}_{\left[ t \right],k}^{\text{H}}+{{{\bm{\tilde{R}}}}_{k}} \right){{{\bm{\hat{h}}}}_{t,k}}}=\\
&N{{N_\text{R}}}{\left( {\frac{1}{{\sqrt {N{N_\text{R}}} }}{{{\bm{\mathord{\buildrel{\lower3pt\hbox{$\scriptscriptstyle\frown$}}
\over h} }}}_{t,k}}} \right)^{\text{H}}}{\left( {{{\bm{I}}_{{{N_\text{R}}}}} \otimes {{{\bm{\hat R}}}_{t,k}}} \right)^{\frac{1}{2}}}\\
&\times\left( {{{{\bm{\hat H}}}_{\left[ t \right],k}}{\bm{\hat H}}_{\left[ t \right],k}^{\text{H}} + {{{\bm{\tilde R}}}_k}} \right){\left( {{{\bm{I}}_{{{N_\text{R}}}}} \otimes {{{\bm{\hat R}}}_{t,k}}} \right)^{\frac{1}{2}}}\left( {\frac{1}{{\sqrt {N{{N_\text{R}}}} }}{{{\bm{\mathord{\buildrel{\lower3pt\hbox{$\scriptscriptstyle\frown$}}
\over h} }}}_{t,k}}} \right).
\end{align*}
By Theorem 3.4 of \cite{Couillet2011Random}, when $N{{N}_\text{R}}\to \infty $, the denominator part is
\begin{align*}
&{\bm{\hat{h}}_{t,k}^{\text{H}}\left( {{{\bm{\hat{H}}}}_{\left[ t \right],k}}\bm{\hat{H}}_{\left[ t \right],k}^{\text{H}}+{{{\bm{\tilde{R}}}}_{k}} \right){{{\bm{\hat{h}}}}_{t,k}}}\xrightarrow{a.s.}\\
&{\text{Tr}}\biggl( {{\left( {{{\bm{I}}_{{{N_\text{R}}}}} \otimes {{{\bm{\hat R}}}_{t,k}}} \right)}^{\frac{1}{2}}}\left( {{{{\bm{\hat H}}}_{\left[ t \right],k}}{\bm{\hat H}}_{\left[ t \right],k}^{\text{H}} + {{{\bm{\tilde R}}}_k}} \right) \biggr.\\
&\times\biggl.{{\left( {{{\bm{I}}_{{{N_\text{R}}}}} \otimes {{{\bm{\hat R}}}_{t,k}}} \right)}^{\frac{1}{2}}} \biggr).
\end{align*}
From the circular property of the trace of the matrix, the above formula is equal to
$${\mathop{\text {Tr}}\nolimits} \left( {\left( {{{\bm{I}}_{{{N_\text{R}}}}} \otimes {{{\bm{\hat R}}}_{t,k}}} \right){{{\bm{\hat H}}}_{\left[ t \right],k}}{\bm{\hat H}}_{\left[ t \right],k}^{\text{H}}} \right) + {\mathop{\text {Tr}}\nolimits} \left( {\left( {{{\bm{I}}_{{{N_\text{R}}}}} \otimes {{{\bm{\hat R}}}_{t,k}}} \right){{{\bm{\tilde R}}}_k}} \right).$$
Use the cyclic properties of the trace on the first term, and by the definition of the trace, the first term is
\begin{align*}
&{\mathop{\text {Tr}}\nolimits} \left( {{\bm{\hat H}}_{\left[ t \right],k}^{\text{H}}\left( {{{\bm{I}}_{{{N_\text{R}}}}} \otimes {{{\bm{\hat R}}}_{t,k}}} \right){{{\bm{\hat H}}}_{\left[ t \right],k}}} \right)=\\
&\sum\limits_{{l_1} = 1,{l_1} \ne t}^{{{N_\text{T}}}} {{\bm{\hat h}}_{{l_1},k}^{\rm H}\left( {{{\bm{I}}_{{{N_\text{R}}}}} \otimes {{{\bm{\hat R}}}_{t,k}}} \right)} {{\bm{\hat h}}_{{l_1},k}},
\end{align*}
and can be further rewritten as
\begin{align*}
N{{N_\text{R}}}\sum\limits_{{l_1} = 1,{l_1} \ne t}^{{{N_\text{T}}}} {{{\left( {\frac{1}{{\sqrt {N{{N_\text{R}}}} }}{{{\bm{\mathord{\buildrel{\lower3pt\hbox{$\scriptscriptstyle\frown$}}
\over h} }}}_{{l_1},k}}} \right)}^{\text{H}}}{{\left( {{{\bm{I}}_{{{N_\text{R}}}}} \otimes {{{\bm{\hat R}}}_{{l_1},k}}} \right)}^{\frac{1}{2}}}}\\
\times{\left( {{{\bm{I}}_{{{N_\text{R}}}}} \otimes {{{\bm{\hat R}}}_{t,k}}} \right){{\left( {{{\bm{I}}_{{{N_\text{R}}}}} \otimes {{{\bm{\hat R}}}_{{l_1},k}}} \right)}^{\frac{1}{2}}}\left( {\frac{1}{{\sqrt {N{{N_\text{R}}}} }}{{{\bm{\mathord{\buildrel{\lower3pt\hbox{$\scriptscriptstyle\frown$}}
\over h} }}}_{{l_1},k}}} \right)}.
\end{align*}
When $N{{N}_\text{R}}\to \infty $, this term will tend to
\begin{align*}
\sum\limits_{{{l}_{1}}=1,{{l}_{1}}\ne t}^{{{N}_{\text{T}}}}{\left( {{\mathbf{I}}_{{{N}_\text{R}}}}\otimes {{{\mathbf{\hat{R}}}}_{{{l}_{1}},k}} \right)\left( {{\mathbf{I}}_{{{N}_\text{R}}}}\otimes {{{\mathbf{\hat{R}}}}_{t,k}} \right)}.
\end{align*}
Considering the characteristics of the Kronecker product with a unit matrix, the denominator is
\begin{equation}\label{eq:eq51}
\begin{aligned}
&{\bm{\hat{h}}_{t,k}^{\text{H}}\left( {{{\bm{\hat{H}}}}_{\left[ t \right],k}}\bm{\hat{H}}_{\left[ t \right],k}^{\text{H}}+{{{\bm{\tilde{R}}}}_{k}} \right){{{\bm{\hat{h}}}}_{t,k}}}\xrightarrow{a.s.}\\
&{{N}_{\text{R}}}\text{Tr}\left( \sum\limits_{{{l}_{1}}=1,{{l}_{1}}\ne t}^{{{N}_{\text{T}}}}{{{{\bm{\hat{R}}}}_{t,k}}{{{\bm{\hat{R}}}}_{{{l}_{1}},k}}+{{{\bm{\hat{R}}}}_{t,k}}\left( \sum\limits_{{{l}_{2}}=1}^{{{N}_{\text{T}}}}{{{{\bm{\tilde{R}}}}_{{{l}_{2}},k}}}+\frac{1}{{{\gamma }_{\text{C}}}}{{\bm{I}}_{N}} \right)} \right).
\end{aligned}
\end{equation}
By substituting (\ref{eq:eq50}) and (\ref{eq:eq51}) into (\ref{eq:eq14}), (\ref{eq:eq16}) can be achieved.

\section{calculation in deterministic equivalent}\label{sec:Appendix_II}
\begin{breakablealgorithm}
\caption{Algorithm to achieve ${\bm{T}_{t,k}}$}
\label{alg:Framwork}
\begin{algorithmic}[1]
\Require Correlation matrix of the equivalent channel corresponding to the $l$-th transmit antenna ${\bm{\Phi }_{l,t,k}}$;
Index of the transmit antenna currently being calculated $t$;
Number of transmit antennas ${{N}_\text{T}}$;
Number of receive antennas ${{N}_\text{R}}$;
Number of repetitions $N$;
Computation accuracy $\Delta$;
\Ensure Matrix ${\bm{T}_{t,k}}$;
\State Initialize ${{e}_{l,t,k}}=0, l=1,\cdots ,{{N}_{\text{T}}}$ and $l\ne t$;
\State Let ${{e}_{l,t,k}}'={{e}_{l,t,k}},l=1,\cdots ,{{N}_{\text{T}}}$ and $l\ne t$;\label{step2}
\State Calculate ${\bm{T}_{t,k}}$ by (\ref{eq:eq23});
\State Calculate ${{e}_{l,t,k}}, l=1,\cdots ,{{N}_{\text{T}}}$ and $l\ne t$ by (\ref{eq:eq25});
\State Calculate the difference between the result of the current iteration and that of the previous iteration $\Delta '=\sum\limits_{l=1,l\ne t}^{{{N}_{\text{T}}}}{{{\left| {{e}_{l,t,k}}-{{e}_{l,t,k}}' \right|}^{2}}}$;
\State if $\Delta '\le \Delta$, return ${\bm{T}_{t,k}}$; otherwise, go to \ref{step2};
\end{algorithmic}
\end{breakablealgorithm}

\section{proof of theorem \ref{THEOREM 2}}\label{sec:Appendix_III}
Following Appendix C of \cite{Zhou2015High}, in the polar coordinate system, the conditional probability density function of the decision variable can be expressed as
\begin{equation}\label{eq:eq52}
p\left( {r,\psi |{s_{t,k}},{{\bm{\bar h}}_{t,k}}} \right) = \frac{r}{{\pi \sigma _{\hat s|{\bm{\bar h}},s,t,k}^2}}\exp \left( { - \frac{{{r^2}}}{{\sigma _{\hat s|{\bm{\bar h}},s,t,k}^2}}} \right).
\end{equation}
The conditional error probability can be calculated as follows:
\begin{align}\label{eq:eq53}
&P\left( {E{\rm{|}}{s_{t,k}},{{\bm{\bar h}}_{t,k}}} \right)\nonumber\\
&=2\int_0^{\pi  - \frac{\pi }{M}} {\int_{{R_{t,k}}\left( \psi  \right)}^\infty  {p\left( {r,\psi |{s_{t,k}},{{\bm{\bar h}}_{t,k}}} \right)drd\psi } }\nonumber\\
&=\frac{1}{\pi }\int_0^{\pi - \frac{\pi }{M}}\exp \bigggl[ - \bm{\hat h}_{t,k}^{\text{H}}\left( {{E_\text{C}}\bm{R}_{\text{W},t,k}^{\text{H}}{\bm{R}_{\text{W},t,k}}} \right){\bm{\hat h}_{t,k}} \bigggr. \nonumber\\
&\times\bigggl. \frac{{{{\sin }^2}\left( {\frac{\pi }{M}} \right)}}{{{{\sin }^2}\left( \theta  \right)}} \bigggr] d\theta,
\end{align}
where
$${R_{t,k}}\left( \psi  \right){\rm{ = }}\frac{{\sin \left( {\frac{\pi }{M}} \right)}}{{\sin \left( {\psi  + \frac{\pi }{M}} \right)}}\left| {{E_\text{C}}\bm{\bar h}_{t,k}^{\text{H}}{\bm{\bar h}_{t,k}}{s_{t,k}}} \right|.$$
Under MPSK modulation, the constellation points are symmetrical, and the probability of each symbol is the same. The conditional error probability averaged over the transmitted symbols is
\begin{equation}\label{eq:eq54}
P\left( {E{\rm{|}}{\bm{\bar h}_{t,k}}} \right){\rm{ = }}\frac{1}{M}\sum\limits_{{s_k} \in S} {P\left( {E{\rm{|}}{s_{t,k}},{\bm{\bar h}_{t,k}}} \right) = } P\left( {E{\rm{|}}{s_{t,k}},{\bm{\bar h}_{t,k}}} \right).
\end{equation}
Considering that the term $\beta {\rm{ = }}\bm{\hat h}_{t,k}^{\text{H}}\left( {{E_\text{C}}\bm{R}_{\text{W},t,k}^{\text{H}}{\bm{R}_{\text{W},t,k}}} \right){\bm{\hat h}_{t,k}}$ is the quadratic form of the zero-mean complex Gaussian random vector ${\bm{\hat{h}}_{t,k}}$ with ${\bm{I}_N} \otimes {\bm{\hat R}_{t,k}}$ as its covariance, the moment generating function of $\beta$ is
\begin{equation}\label{eq:eq55}
\begin{aligned}
&{\Psi _\beta }\left( \mu  \right) = E\left( {{e^{\mu \beta }}} \right) \\
&= {\left[ {\det \left( {{\bm{I}_{N{{N_\text{R}}}}} - \mu \left( {{\bm{I}_N} \otimes {\bm{\hat R}_{t,k}}} \right)\left( {{E_\text{C}}\bm{R}_{\text{W},t,k}^{\text{H}}{\bm{R}_{\text{W},t,k}}} \right)} \right)} \right]^{{\rm{ - }}1}},
\end{aligned}
\end{equation}
where $\mu$ is a dummy variable and the unconditioned average SER ${{\bar{P}}_{\text{e},t,k}}=E\left[ P\left( E\text{ }\!\!|\!\!\text{ }{{\bm{\bar{h}}}_{t,k}} \right) \right]$ calculated using (\ref{eq:eq53}), (\ref{eq:eq54}), (\ref{eq:eq55}), and (\ref{eq:eq35}) can be obtained.

\bibliographystyle{IEEEtran}
\bibliography{access}

\begin{IEEEbiography}[{\includegraphics[width=1in,height=1.25in,clip,keepaspectratio]{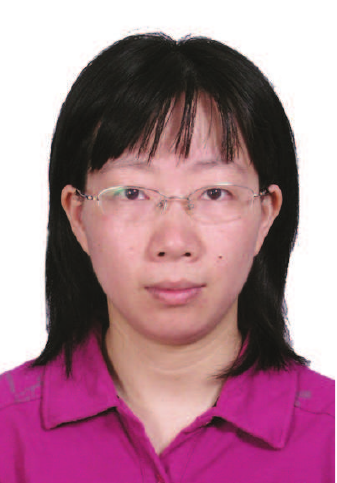}}]{Xiaoyun Hou}
received the B.S. degree and the M.S. degree from Nanjing University of Posts and Telecommunications in
1999 and 2002, and the Ph.D. degree from Shanghai Jiaotong University in 2005. She joined Nanjing University of Posts and Telecommunications, China, in 2005, where she has been an Associate Professor since 2009. She was a visiting scholar at University of California, Davis, CA from 2012 to 2013.
Her current research interests include signal processing techniques in high mobility MIMO communications.
\end{IEEEbiography}

\begin{IEEEbiography}[{\includegraphics[width=1in,height=1.25in,clip,keepaspectratio]{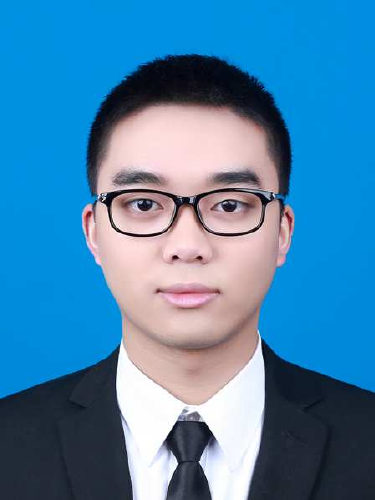}}]{Jie Ling}
received the B.S. degree from Nanjing University of Posts and Telecommunications in 2017, where currently he is working for the M.S. degree. His current research interests include signal processing techniques in high mobility MIMO communications.
\end{IEEEbiography}

\begin{IEEEbiography}[{\includegraphics[width=1in,height=1.25in,clip,keepaspectratio]{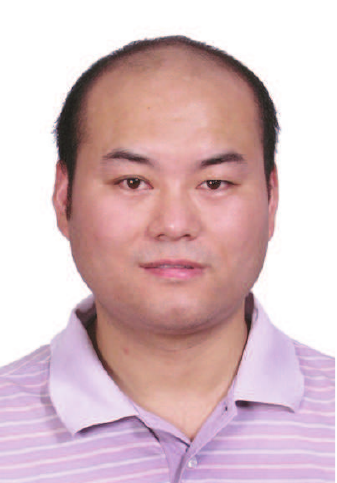}}]{Dongming Wang}
received the B.S. degree from Chongqing University of Posts and Telecommunications in
1999, the M.S. degree from Nanjing University of Posts and
Telecommunications in 2002, and the Ph. D. degree from Southeast
University in 2006. He joined the National Mobile Communications Research Laboratory at Southeast
University, China, in 2006, where he is a Professor now.
He serves as an associate editor for the SCIENCE CHINA Information Sciences.
His current research interests include channel
estimation, distributed antenna systems, and large-scale MIMO systems.
\end{IEEEbiography}

\EOD

\end{document}